\documentclass[authoryear,3p,10pt]{elsarticle}

\usepackage{setspace} 
\usepackage{amsmath,amssymb,mathtools,mathrsfs}
\usepackage[colorlinks=true,citecolor=blue,linkcolor=blue,urlcolor=blue]{hyperref}%
\usepackage[usenames,dvipsnames]{xcolor}

\numberwithin{equation}{section}

\usepackage{color, colortbl}
\definecolor{Gray}{gray}{0.9}
\definecolor{Col1}{rgb}{0.9,1,1} %Kh >> ks is not satisfied
\definecolor{Col2}{rgb}{1,1,0.9}  %Kh >> 1 is not satisfied
\usepackage{graphicx}
\usepackage{natbib}
\usepackage{textcomp}
\usepackage{amstext}
\usepackage{multicol}
\usepackage{tikz}
\usepackage{lettrine}
\allowdisplaybreaks
\usepackage[export]{adjustbox}
\usepackage{ragged2e}
\usepackage{subfig}
\usepackage{float}
\usepackage{booktabs}
\usepackage{makecell, multirow}
\usepackage{tabularx}
\usepackage{siunitx}
\newcommand{\lcurv}{\ell_\mathrm{curv}}
\newcommand{\lvdw}{\ell_\mathrm{vdW}}

\newcommand{\Kh}{\mathcal{K}_{h}}
\newcommand{\Ks}{\mathcal{K}_{s}}

\newcommand{\Sl}{\mathcal{S}}
\newcommand{\Sea}{\mathcal{S}_\mathrm{ea}}
\newcommand{\BPMF}{\pmb{B}_\mathrm{PMF}}
\newcommand{\Bha}{|\tilde {\pmb{B}}|}

\newcommand{\dd}{\mathrm{d}}
\newcommand{\rt}{\Tilde{r}}
\newcommand{\w}{\Tilde{w}}
\newcommand{\Nr}{\Tilde{N}_r}
\newcommand{\Nt}{\Tilde{N}_\theta}
\newcommand{\Upm}{U^+_-}
\newcommand{\Upmt}{\tilde U^+_-}
\newcommand{\psit}{\tilde \psi}
\newcommand{\ut}{\tilde u}
\newcommand{\p}{\mathcal{P}}
\newcommand{\Ksup}{K_\mathrm{sup}}
\newcommand{\rsheet}{\mathcal{R}}
\newcommand{\rin}{L_I}
\newcommand{\rout}{L_O}
\newcommand{\ct}{\tilde C}
\newcommand{\E}{\mathcal{E}}

\newcommand{\red}[1]{{\color{red} #1 }}

\begin{document}

\journal{International Journal of Solids and Structures}

\begin{frontmatter}

\title{\textbf{Two-dimensional crystals on adhesive substrates subjected to uniform transverse pressure}}

\author[add1,add2,add3]{Zhaohe Dai}
\author[add1]{Yifan Rao}
\author[add1]{Nanshu Lu}%\corref{cor1}}
%\ead{nanshulu@utexas.edu}
\address[add1]{Department of Aerospace Engineering and Engineering Mechanics, University of Texas, Austin, TX 78712, United States}
\address[add2]{Mathematical Institute, University of Oxford, Woodstock Rd, Oxford, OX2 6GG, UK}
\address[add3]{Department of Mechanics and Engineering Science, College of Engineering, Peking University, Beijing 100871, China}

\begin{abstract}
In this work we consider bubbles that can form spontaneously when a two-dimensional (2D) crystal is transferred to a substrate with gases or liquids trapped at the crystal-substrate interface. The underlying mechanics may be described by a thin sheet on an adhesive substrate with the trapped fluid applying uniform transverse pressure. What makes this apparently simple problem complex is the rich interplay among geometry, interface, elasticity and instability. Particularly, extensive small-scale experiments have shown that the 2D crystal surrounding a bubble can adhere to and, meanwhile, slide on the substrate. The radially inward sliding causes hoop compression to the 2D crystal which may exploit wrinkling instabilities to relax or partially relax the compression. We present a theoretical model to understand the complex behaviors of even a linearly elastic 2D crystal due to the combination of nonlinear geometry, adhesion, sliding, and wrinkling in bubble systems. We show that this understanding not only successfully predicts the geometry of a spontaneous bubble but also reveals the strain-coupled physics of 2D crystals, e.g.,  the pseudomagnetic fields in graphene bubbles.
\end{abstract}

\end{frontmatter}

\section{Introduction}

As very common scenarios in structural engineering, plates/sheets subjected to lateral loads have been long studied historically \cite[][]{Timoshenko1959}. Recently, interests in this textbook problem have been renewed by the emergence of atomically thin two-dimensional (2D) crystals \cite[][]{Akinwande2017}, particularly their strain engineering---concerning the influence of mechanical strains on the physics of the 2D crystals \cite[][]{Dai2019strain,SANCHEZ2021}. An important mechanics ingredient highlighted by extensive experiments on the strain engineering of 2D materials is that lateral loads can be applied passively by the spontaneous van der Waals (vdW) interactions between the sheet and the substrate \cite[][]{Liechti2019,Dai2020mechanics,Li2021}. Experimental measurements and theoretical prediction of the strain fields in these spontaneous systems are fundamental to the design of strain engineering. The theories, however, are nontrivial due to the complex interplay between the geometrical nonlinearity, the elasticity-adhesion interaction, and the excessive bendability of 2D crystals.

To understand these complexities (particularly under which conditions the theory can be simplified), this work focuses on a relatively simple configuration---axisymmetric \emph{bubbles} \cite[][]{SANCHEZ2021}. When transferring a 2D crystal to a substrate, contamination such as gases and/or liquids are often trapped at the crystal-substrate interface \cite[][]{Frisenda2018recent,Hou2020preparation}. The interfacial vdW adhesion can squeeze the trapped fluids to form bubbles so that the 2D crystal is deformed laterally. Such spontaneously formed bubbles are mostly undesired for 2D crystal based devices because their high performance relies on the flatness of the crystal and the cleanness of the crystal-substrate interface \cite[][]{Kretinin2014electronic}. However, 2D material bubbles have found unique capability in strain engineering that exploits the strain-dependent physics. For example, the coupling between strain and electronic properties in graphene bubbles has been found to produce pseudomagnetic fields (PMFs) of magnitudes on the order of 100 Tesla \cite[][]{Levy2010,Settnes2016}. Such gigantic PMFs might be used for the design of valley filtering and valley splitting devices \cite[][]{Settnes2016}. There are also a number of other exciting examples that have exploited the strain in bubbles such as for the tuning of band gaps \cite[][]{Lloyd2016band,Wang2021out}, piezoelectricity \cite[][]{Ares2020piezoelectricity,Wang2021visualizing}, surface plasma \cite[][]{Fei2016ultraconfined} and friction \cite[][]{Zhang2019tuning} and so on.

The mechanics of the spontaneously formed bubbles under 2D crystals may be described by a thin sheet on an adhesive substrate with the trapped fluids applying a uniform transverse pressure (Fig.~\ref{fig:Schematic}a). A governing mechanism for the mechanical behaviors of 2D crystal bubbles is the elasticity-adhesion (herein referred to as \emph{elastoadhesive}) interactions \cite[][]{Dai2019strain,SANCHEZ2021}. The complexity of this problem comes from the need of a model to describe the elastoadhesive interactions with the consideration of the unique slippery nature of 2D crystals that may allow the sheet to slide on the substrate when subjected to pressure. The radially inward sliding causes hoop compression to the 2D crystal which may exploit wrinkling instability to (partially) relax such compression. Therefore, two more important factors come to interplay with the elastoadhesion of the thin sheet - sliding and wrinkling.\footnote{Note that the residual stress, a.k.a. pretension, in the 2D crystal is one more factor that may further enrich the problem. However, we neglect this effect in this work.}

Experimentally, the crystal-substrate adhesion has been demonstrated to control the geometry (e.g., the aspect ratio) of the spontaneously formed bubbles \cite[][]{Khestanova2016universal,Sanchez2018,Blundo2021experimental,Villarreal2021breakdown}. The sliding of the 2D crystal has been observed using Raman spectroscopy \cite[][]{Kitt2013graphene,Wang2017}. The sliding-induced wrinkling has been found in the suspended or the supported part of the crystal using atomic force microscopy (AFM) \cite[][]{Dai2018interface,Jia2019programmable,Luo2020,Ares2021,Hou2021}. 

A number of theoretical works have addressed the problem of thin sheet bubbles. For example, the adhesion effect in bubbles formed by linearly elastic plates and membranes has been discussed by \cite{Koenig2011}, \cite{Yue2012analytical}, \cite{Wang2013numerical}, and \cite{Boddeti2013mechanics}. An extension for thin hyperelastic sheets has been recently made by \cite{Rao2021elastic}. To carefully interpret the adhesion from experimental results, mixed-mode fracture theory or traction-separation relations have been employed by \cite{Cao2015mixed,Cao2016mixed} and \cite{Wood2017adhesion}. Considering that the surface of 2D materials is atomically smooth, \cite{Khestanova2016universal} and \cite{Sanchez2018} studied the geometry of spontaneously formed bubbles using a vanished sliding resistance at the sheet-substrate interface. 

While these models have recognized sliding and the sliding-caused hoop compression, the interplay between possible wrinkling instabilities \cite[that were observed in for example][]{Ares2021,Dai2021poking} and adhesion remains to be studied. In fact, the wrinkling could play an important role in strain-coupled physics such as PMFs because they could perturb the local strain considerably. However, the task of integrating the elastoadhesion, sliding, and wrinkling into a single model to capture the 2D crystal bubble system is rather nontrivial.

Here, we present a unified model on the mechanics of spontaneously formed thin sheet bubbles with a particular focus on the interplay between nonlinear geometry, elastoadhesion, sliding, and wrinkling. Specifically, we develop a novel "slope discontinuity condition" to characterize the adhesion effect. Inspired by recent works such as \cite{Dai2021poking} and \cite{Davidovitch2021indentation}, we describe the sliding between the sheet and its substrate by a single parameter and the wrinkling of the thin sheet by two parameters (one for the suspended part and the other for the substrate-supported part). We understand the complexity in the apparently simple system by investigating a number of sliding/wrinkling parameter regimes. In each regime, the strain-induced PMFs are discussed to exemplify the potential use of this model in the strain engineering of 2D materials.   

\begin{table}[!htb]
\small
    \caption{\red{A summary of key parameters and variables used in this work.}}
    \label{Parameters}
\begin{center}
\begin{tabular}{ ccl } 
        \midrule
        \midrule
 \textbf{Parameters/Variables} & \textbf{Dimensionless form} & \textbf{Definition}\\
        \midrule
 $a$          &---          & radius of the bubble \\ 
 $h$          &---          & height of the bubble \\ 
 $s$          &---          & sheet-substrate spacing\\
 $r_\mathrm{sheet}$    &$\rsheet=r_\mathrm{sheet}/a$ & physical size of the sheet\\
 $\ell_I$     &$L_I=\ell_I/a$  & size of the inner unwrinkled core\\
 $\ell_O$     &$L_O=\ell_O/a$  & size of the outer wrinkled zone\\
 (Also labeled in Fig. \ref{fig:Schematic})\\
         \cmidrule{1-3}
 $Y$          &---          & Young's modulus $\times$ thickness\\
 $\nu$        &---          & Poisson's ratio\\
 $D$          &---          & bending stiffness\\
 $\tau$       &---          & interfacial shear resistance\\
 $\Gamma$     &$\gamma=\Gamma/Y$ & interfacial adhesion energy\\
          \cmidrule{1-3}
 $\vartheta$  &---          & contact angle (Fig. \ref{fig:Schematic}a)\\
 $\Kh$        &---          & $={Yh^2}/{D}$\\
 $\Ks$        &---          & $={Ys^2}/{D}$\\ 
 $\Sl$        &---          & $={Yh^2}/({\tau a^3})$\\
 $\Sea$       &---          & $={\sqrt{Y\Gamma}}/({\tau a})$\\
          \cmidrule{1-3}
 $r$          &$\rt={r}/{a}$& polar coordinate\\
 $\theta$     &---          & polar coordinate\\
 $u(r)$       &$\ut=u/a$    & in-plane displacement\\
 $w(r)$       &$\w=w/a$     & out-of-plane displacement\\
 $\epsilon_r$ &---          & radial strain\\
 $\epsilon_\theta$&---      & hoop strain\\
 $N_r$        &$\Nr=N_r/Y$  & radial stress resultant\\
 $N_\theta$   &$\Nt=N_\theta/Y$  & hoop stress resultant\\
 $p$          &$\p=pa/Y$    & transverse pressure\\
 $\psi$       &$\psit=\psi/(Ya)$ & Airy stress function \\
         \midrule
         \midrule
\end{tabular}
\end{center}
\end{table}

\section{The model\label{sec:model}}

We begin by introducing equations and parameters that we will use in the latter sections, including the simplified strain-PMF relation in graphene, reduced Föppl–von Kármán equations for thin sheet elasticity, slope discontinuity conditions due to the presence of adhesion, and some useful parameters to quantify the ability of the sheet to slide and to wrinkle on an adhesive substrate.  

\subsection{Pseudomagnetic fields}
The specific mechanics-physics coupling example in 2D materials considered in this paper is the strain-gradient-induced PMFs. Briefly, in-plane strains can introduce an effective gauge field
\begin{equation}
    \pmb{A}=A_\mathrm{PMF}(\epsilon_{xx}-\epsilon_{yy},-2\epsilon_{xy})
\end{equation} in the low-energy Dirac Hamiltonian, where the coupling constant ${A}_\mathrm{PMF}\approx 7 \mathrm{~\mu m\,T}$ can be further related to hopping energy, Fermi velocity, and electron charge \cite[][]{Guinea2010}. This gauge field shifts the Dirac cones of graphene at points K and K’ in the opposite directions, reminiscent of the effect of a perpendicularly applied magnetic field \cite[][]{Guinea2010}. Such PMFs can be related to the strain-caused gauge field by
\begin{equation}
    \BPMF=\pmb{\nabla}\times \pmb{A}.
\end{equation} Note that this unique coupling has raised outstanding inverse problems regarding how to design strain fields to achieve deterministic PMFs \cite[][]{Guinea2010,Zhu2015,Akinwande2017,Hu2019PMF}. For the primary interest of this work --- an axisymmetric graphene bubble of radius $a$ , the magnitude of the generated PMFs can be expressed as \cite[][]{Klimov2012,Zhu2014}
\begin{equation}
    |\BPMF|={A_\mathrm{PMF}}\sin{3\theta}\left[\frac{2(\epsilon_r-\epsilon_\theta)}{r}-\frac{\dd (\epsilon_r-\epsilon_\theta)}{\dd r}\right].
    \label{eq:BPMF1}
\end{equation}
Elementary geometry of bubbles suggests that both strain components are proportional to the square of the bubble's aspect ratio, i.e. $\epsilon\sim h^2/a^2$; Therefore, a rescaled magnitude of PMFs
\begin{equation}
    \Bha=\frac{|\BPMF|a^3}{A_\mathrm{PMF}h^2}
    \label{eq:BPMF2}
\end{equation}
will be used in the sequel to illustrate the effects of sliding and wrinkling on the distribution of PMFs. We will also show that the adhesion effect is mainly reflected by the characteristic aspect ratio of the bubble and, therefore, the maximum of $|\BPMF|$.

\begin{figure}[t]
    \centering
    \vspace{0cm}
    \includegraphics[width=16cm]{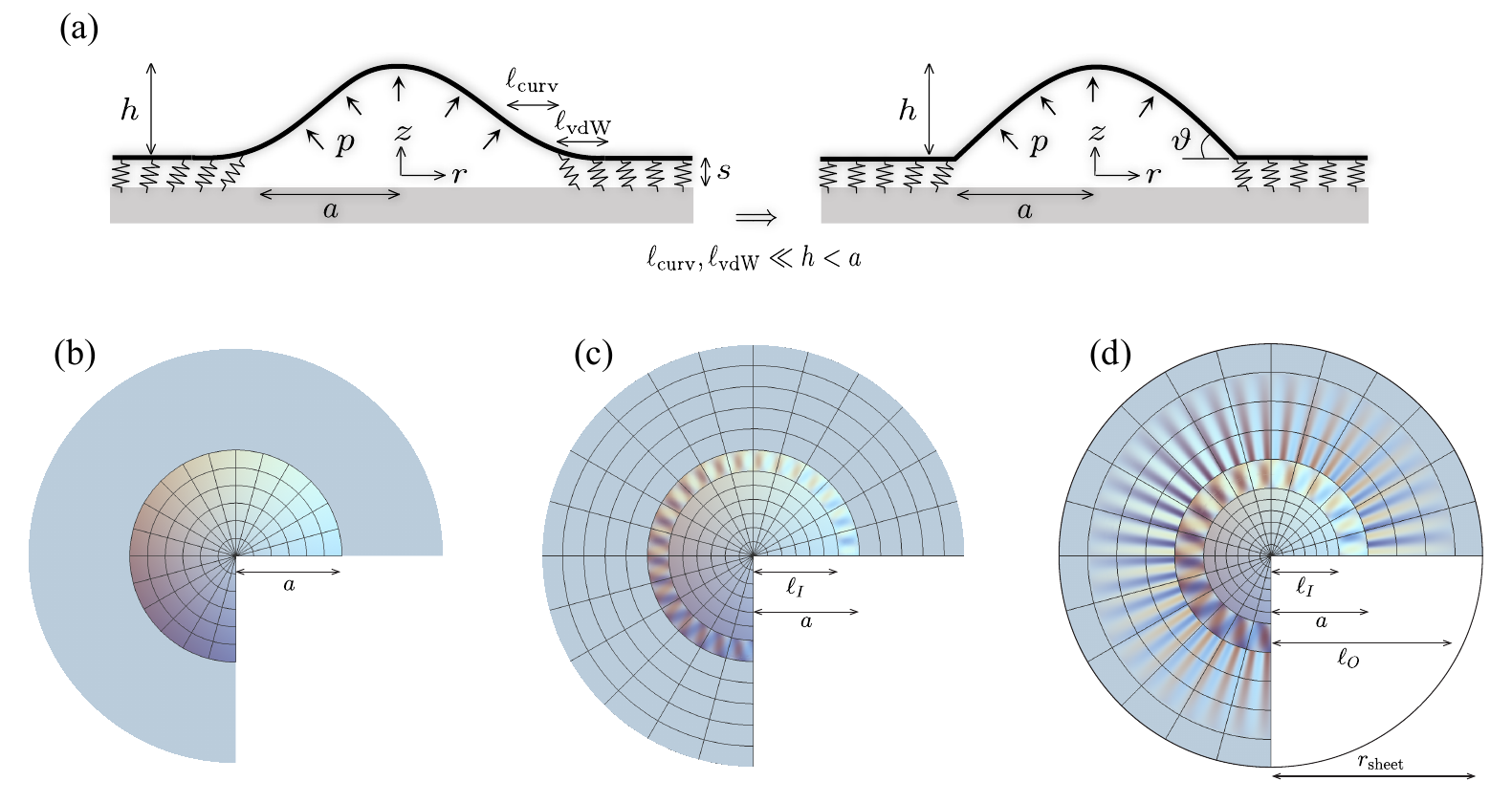}
    \caption{Schematic illustration and notation for the analysis of 2D crystal bubbles. (a) Left: A generic scenario with the consideration of the bending effect of the thin sheet and the finite `process zone' near the edge of delamination due to the sheet-substrate vdW interactions. We describe the tangential sheet-substrate interactions by a uniform, constant shear stress and the normal interactions by a linear, elastic Winkler foundation (with an initial thickness of $s$). Right: This paper focuses on a class of highly bendable thin sheets - 2D crystals. We consider vanishing/infinite shear stresses associated with the radial, inward slippage of the sheet and finite adhesion for the foundation to break (but the size of the process zone is small compared with the bubble size). As a result, the local curvature of the sheet diverges at the edge of the bubble and a non-zero local slope could be observed with the value determined by adhesion competing against elasticity. (b) No sliding: an infinite interfacial shear stress is considered so that the in-plane displacement is fixed at the edge of the bubble and the strain fields are trivial in the whole supported region. (c) Sliding and wrinkling only in the suspended region: the sheet is allowed to slide on the substrate without any tangential resistance (so a wrinkling zone $[\ell_I,a]$ occurs in the suspended region); But the normal sheet-substrate interactions are assumed to be substantial enough to suppress the formation of wrinkling in the supported region. (d) Sliding and wrinkling in both suspended and supported regions $[\ell_I,\ell_O]$: vanishing tangential interactions and finite normal interactions are considered at the sheet-substrate interface so that the sheet is allowed to slide as well as wrinkle in the supported region. In this case, the stresses decay very slowly ($\sim r^{-1}$) so that the boundary conditions at the outer edge of the thin sheet (even with $r_\mathrm{sheet}\gg a$) become important. The condition of zero displacement at $r=r_\mathrm{sheet}$ is used in this work.}
    \label{fig:Schematic}
\end{figure}

\subsection{Equilibrium equations}
To further relate the radial and hoop strains to the in-plane displacement $u(r)$ and the out-of-plane displacements $w(r)$ in a circular bubble, we assume moderate rotation:
\begin{equation}
    \epsilon_r=\frac{\dd u}{\dd r}+\frac{1}{2}\left(\frac{\dd w}{\dd r}\right)^2\quad\text{and}\quad\epsilon_\theta=\frac{u}{r}.
    \label{eq:Kinematics}
\end{equation}
With Hooke's law, the corresponding stress resultants are calculated to be
\begin{equation}
    N_r=\frac{Y}{1-\nu^2}(\epsilon_r+\nu \epsilon_\theta)\quad\text{and}\quad N_\theta =\frac{Y}{1-\nu^2}(\epsilon_\theta + \nu \epsilon_r),
    \label{eq:Hooke}
\end{equation} where $\nu$ is the Poisson's ratio of the sheet and $Y$ is defined by Young's modulus times `nominal' thickness of the sheet (which is often called in-plane stretching stiffness of the sheet with unit of $\mathrm{N/m}$ instead of $\mathrm{N/m^2}$).

In the presence of a uniform transverse pressure $p$ and absence of horizontal shear loads, the equilibrium equations read, according to Föppl–von Kármán (FvK) equations,
\begin{gather}
    D\nabla^4w - N_r\kappa_r-N_\theta\kappa_\theta - p=0\label{eq:FvKEqns1}\\
    {\dd}(rN_r)/{\dd r}-N_\theta=0,    
    \label{eq:FvKEqns2}
\end{gather}
where $D$ is the bending stiffness of the sheet. The radial and circumferential curvatures approximate
\begin{equation}
    \kappa_r\approx\frac{\dd^2 w}{\dd r^2}\quad\text{end}\quad\kappa_\theta\approx\frac{1}{r}\frac{\dd w}{\dd r},
\end{equation}
under the assumption of moderate rotation. Note that the application of FvK equations to monolayer 2D crystal should adopt a $Y$--independent bending stiffness \cite[][]{Wang2013numerical,Ahmadpoor2017thermal}.

The bending effect gives rise to a length scale $\lcurv\sim(D/Y)^{1/2}$ over which the sheet curves with a finite curvature near the edge of the bubble and connects to the supported region (Fig.~\ref{fig:Schematic}a). The radius of this local curvature would be comparable to the elastoadhesive length scale $(D/\Gamma)^{1/2}$ when the adhesion effect is considered \cite[][]{Majidi2009,Bico2018}. For 2D materials, however, the significant contrast between small bending stiffness and large stretching stiffness typically leads to $\lcurv \ll a$ \cite[][]{Zhang2011bending,Lu2009elastic,Wang2019bending,Han2020}. We may assume that the bending effect is not important in the mechanics of the suspended region so that the out-of-plane equilibrium equation \eqref{eq:FvKEqns1} can be simplified by neglecting the bending term.\footnote{At this moment we are discussing ``regular'' FvK equations. Later, we will discuss the importance of bending soon when wrinkling instabilities occur and modify the compatibility of regular FvK equations.} A more appropriate parameter that justifies this simplification can be given by comparing the typical bending energy $\sim D\kappa^2\sim Dh^2/a^4$ to the stretching energy $\sim Y\epsilon^2\sim Yh^4/a^4$ associated with the bubble of height $h$ and radius $a$. In fact, the neglect of bending energy needs $\lcurv\ll h$, i.e.,
\begin{equation}
    \Kh=\frac{Yh^2}{D}\gg1,
    \label{eq:FvKh}
\end{equation}
which is an FvK number (by using the bubble height as the length scale).

This paper will focus on 2D crystals whose $\Kh\gg1$. We then rewrite the first FvK equation \eqref{eq:FvKEqns1}: 
\begin{equation}
    \psi\frac{\dd w}{\dd r}+\frac{pr^2}{2}=0,
    \label{eq:Membrane1}
\end{equation}
where we integrated once (the integration constant vanishes due to the symmetry, i.e., $\dd w/\dd r=0$ at $r=0$); we also used Airy stress function $\psi$ and expressed stresses by
\begin{equation}
    N_r=\frac{\psi}{r}\quad\text{and}\quad N_\theta=\frac{\dd \psi}{\dd r}
\end{equation} so that the second FvK equation \eqref{eq:FvKEqns2} is satisfied automatically. The equation for $\psi$ is given by the compatibility condition \cite[][]{Mansfield1989}:
\begin{equation}
    \frac{Y}{2}\left(\frac{\dd w}{\dd r}\right)^2 + r\frac{\dd}{\dd r}\left[\frac{1}{r}\frac{\dd}{\dd r}\left(r\psi\right)\right]=0.
    \label{eq:Membrane2}
\end{equation}
When the sheet is not allowed to slide on its substrate, three \emph{no-sliding} boundary conditions arise naturally:
\begin{equation}
    u(0)=\left.{\frac{1}{Y}\left(r\frac{\dd \psi}{\dd r}-\nu \psi\right)}\right|_{r=0}=0,~u(a)=\left.{\frac{1}{Y}\left(r\frac{\dd \psi}{\dd r}-\nu \psi\right)}\right|_{r=a}=0,~w(a)=0,
    \label{eq:BC_CaseI}
\end{equation}
denoting zero in-plane displacements at the center and the edge of the bubble, and zero deflection at the edge, respectively.

Before discussing the last boundary condition that selects a specific pressure $p$, we introduce the non-dimensionalization used throughout this work. In experiments, the radius is often measurable and the stretching modulus of 2D materials is known. We therefore use them to rescale the system:
\begin{equation}
    \Nr=\frac{N_r}{Y},~\Nt=\frac{N_\theta}{Y},~\psit=\frac{\psi}{Ya},~\ut=\frac{u}{a},~\rt=\frac{r}{a},~\w=\frac{w}{a},\p=\frac{pa}{Y}.
    \label{eq:NonD}
    %~L_I=\frac{\ell_I}{a},~L_O=\frac{\ell_O}{a},~R_s=\frac{r_\mathrm{sheet}}{a}.
\end{equation}
Note that, however, when the radius of the bubble is not known \emph{a priori}, the volume of the substance trapped within the bubble provides a length scale that plays the role of $a$ in the re-scaling.  

\subsection{$\mathrm{Griffith/JKR}$-type adhesion}
How is the pressure in \eqref{eq:Membrane1} selected by the competition between adhesion and elasticity? To answer this question, a discussion about the size of the ``cohesive/process zone''---across which the interface energies changes---is necessary. 

\emph{Length scale for the process zone.} The mechanics of 2D material interfaces are rather complex at the scale of a few nm \cite[][]{Zhang2018structural,Xue2022}. For bubble systems with radii of tens of nm or larger, however, a simple model might be adopted: The tangential sheet-substrate interactions are represented by a shear stress $\tau$ \cite[][]{Jiang2014interfacial,Dai2016mechanical,Wang2017} and the normal sheet-substrate interactions are represented by an array of linear springs of constant stiffness $K_\mathrm{sup}$ and initial thickness $s$ (Fig.~\ref{fig:Schematic}a) \cite[][]{Ares2021}. The adhesion energy between the sheet and the substrate $\Gamma$ defines the accumulated energy for these vertical springs within a unit area to break. We then have
\begin{equation}
    \Ksup\sim\Gamma/s^2.
    \label{eq:Ksup}
\end{equation}
Note that we have neglected the variation in the adhesion energy due to the mixed normal and tangential deformation of the interface \cite[or mode mixity as discussed in][]{Liechti2019,Dai2020mechanics}. 

This simple model for the adhering part of the thin sheet is a combination of the shear-lag model with uniform shear stress and the Winkler foundation model with a constant spring stiffness. The vdW process zone involves vertical deformation of the interface so its typical length $\lvdw$ (as shown in Fig. \ref{fig:Schematic}a) is mainly controlled by the foundation model. The detailed $\lvdw$, of course, depends on whether the sheet is in bending or stretching mode (or how the sheet slides) in the supported region, which is not as clear as the situation in the suspended region. That bending mode domination gives a typical Winkler length scale: $$\lvdw^\mathrm{b}\sim (D/\Ksup)^{1/4}.$$ Otherwise, the stretching-mode length scale is $$\lvdw^\mathrm{s}\sim(N_r/\Ksup)^{1/2}\sim(Yh^2/\Ksup a^2)^{1/2}.$$ However, this work avoids this uncertainty by focusing on systems with
\begin{equation}
    \lvdw=\max\left\{{\lvdw^\mathrm{b},\lvdw^\mathrm{s}}\right\}\ll a,
\end{equation} which merely requires $s\ll h$ or
\begin{equation}
    \Ks=\frac{Ys^2}{D}\ll\Kh,
    \label{eq:FvKs}
\end{equation}
where we used \eqref{eq:FvKh} and $\Gamma/Y\sim h^4/a^4$ --- a self-consistent conclusion we shall draw shortly from both simple analysis \eqref{eq:gammaLaw} and detailed numerical results (Fig.~\ref{fig:gammaLaw}). 

\emph{Adhesive boundary conditions.} The small process zone assumption allows the interface to be exclusively characterized by its adhesion energy $\Gamma$, which is identical to the assumption used in the Griffith's theory of fracture and the JKR theory of adhesion \cite[][]{Griffith1921,Johnson1971surface}. We then determine $p$ in \eqref{eq:Membrane1} following a similar idea. The first step is to calculate the total energy of the bubble system:
\begin{equation}
    \Pi=U_\mathrm{elastic}-pV+\pi a^2\Gamma,
    \label{eq:PotentialEnergy}
\end{equation}
where $U_\mathrm{elastic}$ is the sum of elastic strain energy in the thin sheet and $V$ is the volume of the bubble.\footnote{Using $pV$ as the potential
energy of external forces has implied the incompressibility of the trapped substance in the bubble so $p$ is a Lagrange multiplier. When the substance is compressible, a specific $p-V$ law is required to provide a modified version of potential
energy \cite[for example, the ideal gas law used in ][]{Boddeti2013mechanics}. However, we expect the variation method to produce the same equilibrium equations and slope-discontinuity conditions as we presented in the main text. Change in compressibility would only vary the final volume of the bubble with the number of trapped molecules fixed.} However, exact solutions to $U_\mathrm{elastic}$ in this nonlinear problem are elusive in general. The standard numerical method needs to minimize \eqref{eq:PotentialEnergy} with respect to the bubble radius (i.e. $\partial\Pi/\partial a$=0). However, the minimization procedure with an additional constraint on the bubble volume is very tedious, particularly when complex sliding and wrinkling (to be introduced) come into play. Here, inspired by the wetting problem of a drop on a substrate \cite[][]{Rao2021elastic}, we use the variational method with a ``no-pinning'' condition:
\begin{equation}
    \delta\Pi=0\quad\text{with}\quad\delta a\neq0.
    \label{eq:VariationMethod}
\end{equation}

The detailed derivation can be found in \ref{Sec:Appendix}. We point out two main results of the variational analysis: i) The equilibrium equations \eqref{eq:Membrane1} and \eqref{eq:Membrane2} can be reproduced using an appropriate form of elastic strain energy density; ii) The inclusion of adhesion is equivalent to specifying a discontinuous slope of the thin sheet across the edge of the bubble or a \emph{contact angle} at the \emph{contact line} (see the left panel of Fig. \ref{fig:Schematic}a):
\begin{equation}
    \cos\vartheta=\frac{N_r^--\left(\Gamma-\Upm\right)}{N_r^-},
    \label{eq:ContactAngle}
\end{equation}
where 
\begin{equation}
    \Upm=\left.\left(\tfrac{1}{2}N_\theta\epsilon_\theta-\tfrac{1}{2}N_r\epsilon_r\right)\right|^{r=a^+}_{r=a^-}
    \label{eq:EnergyJump}
\end{equation}
represents certain ``energy jump'' across the contact line. We will use specific cases in later sections to understand this quantity. We note that this slope discontinuity condition is conceptually similar to the moment/curvature discontinuity condition reported in the adhesion problems of plates \cite[where the bending effect can smooth out the deflection across a contact line,][]{Majidi2009}. 

We emphasize three properties of \eqref{eq:ContactAngle} here: i) \eqref{eq:ContactAngle} appears very generic, i.e., applicable regardless of sliding and wrinkling (see \ref{Sec:Appendix}). ii) $\Upm\neq0$ in general due to possible ``phase'' changes, such as the thin sheet from wrinkled to unwrinkled, the sheet-substrate interface from attached to detached across the contact line, and so on; iii) Since $\epsilon\sim h^2/a^2$, $N_r^-\sim Yh^2/a$, $\cos\vartheta\sim 1-h^2/a^2$, $\Upm\sim Yh^4/a^4$, \eqref{eq:ContactAngle} immediately suggests
\begin{equation}
    \gamma=\Gamma/Y\sim h^4/a^4,
    \label{eq:gammaLaw}
\end{equation}
where $\gamma$ characterizes the strength of vdW adhesion with respect to the stretching stiffness of the sheet. Consistent with previous work \cite[][]{Khestanova2016universal,Sanchez2018}, \eqref{eq:gammaLaw} suggests that $\gamma$ determines the deformation of the sheet in a spontaneous system and thus selects a specific pressure acting on the thin sheet.

\subsection{The sliding number\label{sec:SlidingAbility}}
The analysis in the preceding sections makes the adhesion problem of \emph{no-sliding} bubbles very simple, which is to solve the ODEs \eqref{eq:Membrane1}
and \eqref{eq:Membrane2} subjected to the no-sliding boundary conditions given in \eqref{eq:BC_CaseI} and a slope-jump condition provided by \eqref{eq:ContactAngle}. However, due to their smooth surfaces, 2D materials such as graphene can easily slide on other substrates \cite[][]{Hod2018,Dai2020mechanics}, breaking down \eqref{eq:BC_CaseI}. To describe the sliding ability of the sheet on a substrate we need to go back to the shear-lag model assuming uniform shear stress $\tau$ at the sheet-substrate interface. In particular, using the same model, a recent work by \cite{Dai2021poking} has identified a single  parameter related to sliding called the sliding number:
\begin{equation}
    \Sl=\frac{Yh^2}{\tau a^3}
\end{equation}
by comparing the driving force (i.e. the radial, inward membrane force $\sim N_ra\sim Yh^2/a$) and the resisting force (i.e. the net outward frictional force $\sim\tau a^2$). Importantly, it was found that the interface can be treated as no-sliding as $\Sl\ll1$ and no-friction as $\Sl\gg1$ \cite[][]{Dai2021poking}. The two opposite limits are of the primary interests of this work. In addition, we modify the sliding number slightly by combining it with \eqref{eq:gammaLaw}:
\begin{equation}
    \Sea=\frac{\sqrt{Y\Gamma}}{\tau a},
    \label{eq:SlidingParameter}
\end{equation}
since the driving force for the sliding in the spontaneous systems comes essentially from the elasto-adhesive interaction.

\subsection{Two wrinkling parameters\label{sec:WrinklingAbility}}

A direct outcome of the radially inward sliding is the in-plane hoop compression to the thin sheet \cite[][]{Davidovitch2011prototypical,Box2019dynamics,Dai2020Radial}. It is natural to think of wrinkling by which an ultrathin sheet in the suspended region releases the compression \cite[][]{Vella2019}. In the supported region, however, the wrinkling ability of the thin sheet requires some examination. For this purpose, we adopt the model presented in \cite{Davidovitch2021indentation} to determine under which condition the thin sheet wrinkles and how much the wrinkling would release the compression. 

The basic idea is to compare the hoop stresses in the thin sheet by allowing and forbidding the wrinkling \cite[][]{Davidovitch2021indentation}: if the sheet is forced to maintain the compression, it has to endure the ``bare'' compressive stress
\begin{equation}
    N_\theta^{bare}=-\alpha Y(h/a)^2,
    \label{eq:N_bare}
\end{equation} where $\alpha$ is a prefactor that scales as $O(1)$ for $r<a$ and decays with the increasing $r$ for $r>a$. If the sheet is allowed to wrinkle to release the compressive stress and the hoop arc-length is nearly inextensible, the hoop stress can be released to a residual value \cite[][]{Davidovitch2021indentation}:
\begin{equation}
    N_\theta^{res}\sim-2D/\lambda^2,
    \label{eq:N_res}
\end{equation}
where $\lambda$ is the wavelength of the radial wrinkles. 

Under the constraint of inextensional arc-length, the bending of the thin sheet favors large $\lambda$ (small curvatures) while both radial membrane tension and the Winkler springs favor small $\lambda$ (thus small amplitudes). In particular, a ``$\lambda$ law'' was reported by \cite{Cerda2003geometry} and generalized by \cite{Paulsen2016curvature}:
\begin{equation}
    \lambda\sim2\pi(D/K_\mathrm{eff})^{1/4}.
    \label{eq:lambdaLaw}
\end{equation}
In the bubble problem, $K_\mathrm{eff}\sim N_r/r^2\sim Yh^2/a^4$ in the suspended region while $K_\mathrm{eff}\sim \Ksup$ in the supported region as appreciated in \cite{Ares2021}. Therefore, \eqref{eq:N_res} can be rewritten as:
\begin{equation}
    N_\theta^{res}\sim- Y\left(\frac{h}{a}\right)^2\times\begin{cases}
    \Kh^{-1/2},\quad r/a<1\\
    \Ks^{-1/2},\quad r/a>1.
    \end{cases}
    \label{eq:N_res2}
\end{equation}
Note that to derive \eqref{eq:N_res2} we have used the definition in \eqref{eq:FvKh} and \eqref{eq:FvKs} and the scaling in \eqref{eq:Ksup} and \eqref{eq:gammaLaw}. 

By comparing the residual stress \eqref{eq:N_res2} and the bare stress \eqref{eq:N_bare} we have: \textit{In the suspended region}, $\alpha\sim O(1)$ so the high bending/wrinkling ability $(\Kh\gg1)$ guarantees $N_\theta^{res}\ll N_\theta^{bare}$. In other words, the compressive hoop stress, if any, will be largely relieved by the formation of wrinkling. \textit{In the supported region}, however, $\alpha\to0$ as $r\to\infty$. Whether the wrinkling instability occurs in a position depends on not only the detailed wrinkling/bending ability of the sheet $\Ks$ in the supported region but also how far this position is away from the bubble. We then study such rich behavior of spontaneous bubbles caused by the interplay of sliding, wrinkling, and adhesion by focusing on several specific $\Sea$ (the sliding ability of sheet on the substrate), $\Kh$ (the wrinkling ability of the sheet inside the bubble), and $\Ks$ (the wrinkling ability of the sheet outside the bubble). Meanwhile, the corresponding strain fields and PMFs in each parameter regime will be discussed. 

\begin{table}[!htb]
\small
    \caption{A summary of numerical results for the mechanical and pseudomagnetic behavior of a graphene sheet with $\nu=0.165$ in different $\left\{\Sea,\Kh,\Ks\right\}$ regimes. Note that all of the regimes require $\Kh\gg1$ to ensure negligible bending effect in the suspended region. $\dag$ in the second last row denotes that the parameter weakly depends on $\rsheet$ and its value is calculated using $\rsheet=100$. $\ddag$ in the last row denotes the case considering the detailed residual stress in wrinkled, supported region. The parameters are not provided because they depend on both $\rsheet$ and $\Ks$ and should lie in between those in the case $\Ks\ll1$ (the third last row) and $\Ks\gg\rsheet^2$ (the second last row).}
    \label{tab}
    \centering
        \begin{tabular}{ ccccccccc }
        \midrule        \midrule
        \multicolumn{3}{c}{{Regime}}
                    & {Asp. ratio}
                    & {Max. strain}
                 &\multicolumn{2}{c}{\makecell{{Max.} $\Bha$}}
                                & \multicolumn{2}{c}{\makecell{{Wrinkled zone}}} \\
        \cmidrule{1-3}
        \cmidrule{6-7}
        %\cmidrule{8-9}
        {Case} & {Sliding} & {Wrinkling}& {$\frac{h}{a}/\gamma^{1/4}$} & {$\epsilon(0)/\left(\frac{h}{a}\right)^2$}  & {Abs. Value} & {Position} & {$\ell_I/a$} & {$\ell_O/a$}\\
        \midrule
        No sliding   & {$\Sea\ll1$}  & {$\Ks\ll\Kh^2$}  &  0.84 & 0.74  & 0.62  & {$r=a^-$} & --- & ---\\
        \cmidrule{1-9}
        {\multirow{3}{*}{\makecell{Sliding \& \\ wrinkling in\\ $[\ell_I,a]$}}} & & & & & & & & \\
        & {$\Sea\gg1$}  & {$\Ks\ll1$}  & 0.98  & 0.42  & 2.87  & {$r=a^+$} & 0.86 & ---\\
         & & & & & & & &\\
        \cmidrule{1-9}
        {\multirow{3}{*}{\makecell{Sliding \& \\ wrinkling in\\ $[\ell_I,\ell_O]$}}} & & & & & & & &\\
        & {$\Sea\gg1$}  & {$\Ks\gg\rsheet^2$} & 1.19$^{\dag}$  & 0.22$^{\dag}$  & 0.92$^{\dag}$  & {$r=\ell_I^+$}  & 0.71$^{\dag}$ & {$\sqrt{\frac{1-\nu}{1+\nu}}\rsheet$} \\
         & & {$\Ks\ll\Kh$}& & & & & & \\
        \cmidrule{1-9}
        {\multirow{3}{*}{\makecell{Sliding \& \\ wrinkling in\\ $[\ell_I,\ell_O]$}}} & & & & & & & &\\
        & {$\Sea\gg1$}  & {$1\ll\Ks\lesssim\rsheet^2$}  & {${\ddag}$}  & {${\ddag}$} & {${\ddag}$} & {$\ell_I^+$ or $a^+$}  & {${\ddag}$} & {$\ll \rsheet$}\\
         & &{$\Ks\ll\Kh$}& & & & & &\\
        \midrule        \midrule
        \end{tabular}
\end{table}

\begin{figure}[!htp]
    \centering
    \vspace{0.0cm}
    \includegraphics[width=9cm]{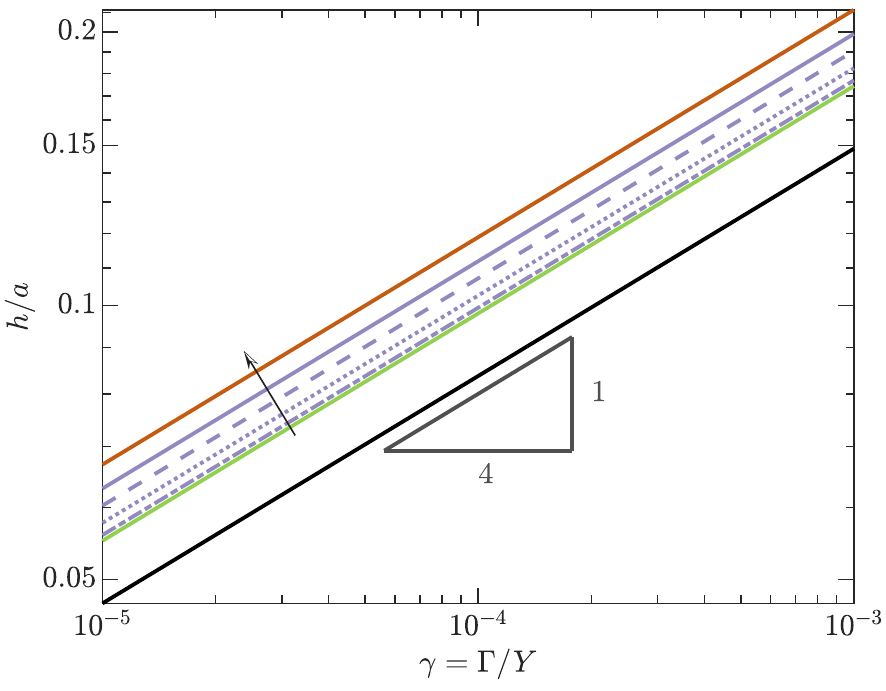}
    \caption{Aspect ratio $h/a$ of the deformed thin sheet as a function of the strength of the vdW adhesion $\Gamma$ with respect to the stretching stiffness $Y$ of the sheet for various cases: no-sliding (black), sliding but wrinkling only in the suspended region (green), sliding and wrinkling in both suspended and supported regions (purple and red). The red curve assumes a vanished residual hoop stress or a large wrinkling ability of the thin sheet ($\Ks\gg\rsheet^2=r_\mathrm{sheet}^2/a^2$) in the wrinkled, supported region while the purple curves considers small but non-zero residual stresses or a moderately large wrinkling ability  ($1\ll\Ks\lesssim\rsheet^2$). Specifically, from dash-dotted, dotted, dashed, to solid curves (along with the direction of the black arrow), the wrinkling ability of the sheet increases from $\rsheet^2/100$,  $\rsheet^2/10$, $\rsheet^2$, to $10\rsheet^2$, respectively. In this work, $\rsheet=100$ is used.}
    \label{fig:gammaLaw}
\end{figure}

\section{No sliding\label{sec:NoSliding}}
\subsection{Regime}
$$\Sea\ll1,\quad\Kh\gg1,\quad\Ks\ll\Kh^2$$

We begin with the simplest case---no sliding (Fig.~\ref{fig:Schematic}b). This case requires a strong shear stress at the sheet-substrate interface, i.e. $\Sea\ll1$ \eqref{eq:SlidingParameter}. Besides, we focus on vanished bending effect in the suspended region by requiring a large FvK number $(\Kh\gg1)$. The strong shear stress would cause the membrane tension to die out quickly in the supported region. Consequently, the small process zone condition \eqref{eq:FvKs} in this case can be slightly released by only requiring $\lvdw^\mathrm{b}\ll a$ or $\Ks\ll\Kh^2$. 

\subsection{Theory}
As discussed in the preceding sections, the problem is to solve the equilibrium equation \eqref{eq:Membrane1}
and compatibility condition \eqref{eq:Membrane2} subjected to the no-sliding boundary conditions \eqref{eq:BC_CaseI} and the slope-jump condition \eqref{eq:ContactAngle}. Following the nondimensionalization given in \eqref{eq:NonD}, the rescaled form of this boundary value problem reads
\begin{equation}
    \psit\frac{\dd \w}{\dd \rt}+\frac{1}{2}\p\rt^2=0,
    \label{eq:Membrane1Rescaled}
\end{equation}
and 
\begin{equation}
    \frac{1}{2}\left(\frac{\dd \w}{\dd \rt}\right)^2 + \rt\frac{\dd}{\dd \rt}\left[\frac{1}{\rt}\frac{\dd}{\dd \rt}\left(\rt\psit\right)\right]=0.
    \label{eq:Membrane2Rescaled}
\end{equation}
subjected to
\begin{equation}
    \lim_{\rt\to0}\left(\rt\psit'-\nu \psit\right)=\psit'(1)-\nu \psit(1)=\w(1)=0.
    \label{eq:BC_CaseI_Rescale}
\end{equation}
To solve for the unknown $\p$ we note that the the dimensionless form of the energy jump \eqref{eq:EnergyJump} is $$\Upmt=\Nr(1)\epsilon_r(1)/2,$$ with which the local contact angle \eqref{eq:ContactAngle} satisfies
\begin{equation}
   \Nr(1-\cos\vartheta)=\gamma-\tfrac{1}{2}\Nr\epsilon_r,
    \label{eq:ContactAngleI}
\end{equation} at $\rt=1$.
It is also worth noting that \eqref{eq:ContactAngleI} is an axisymmetric version of the crack propagation criterion in peeling tests with the radial membrane tension playing the role of the peeling force \cite[see equation (2) in][]{Kendall1975}. Though often neglected in peeling tests because of the small strain, the elastic term $\Upmt$ is important here as the peeling/contact angle $\vartheta$ is small. 

\subsection{Numerical results}

\begin{figure}[!t]
    \centering
    \vspace{0.0cm}
    \includegraphics[width=14.5cm]{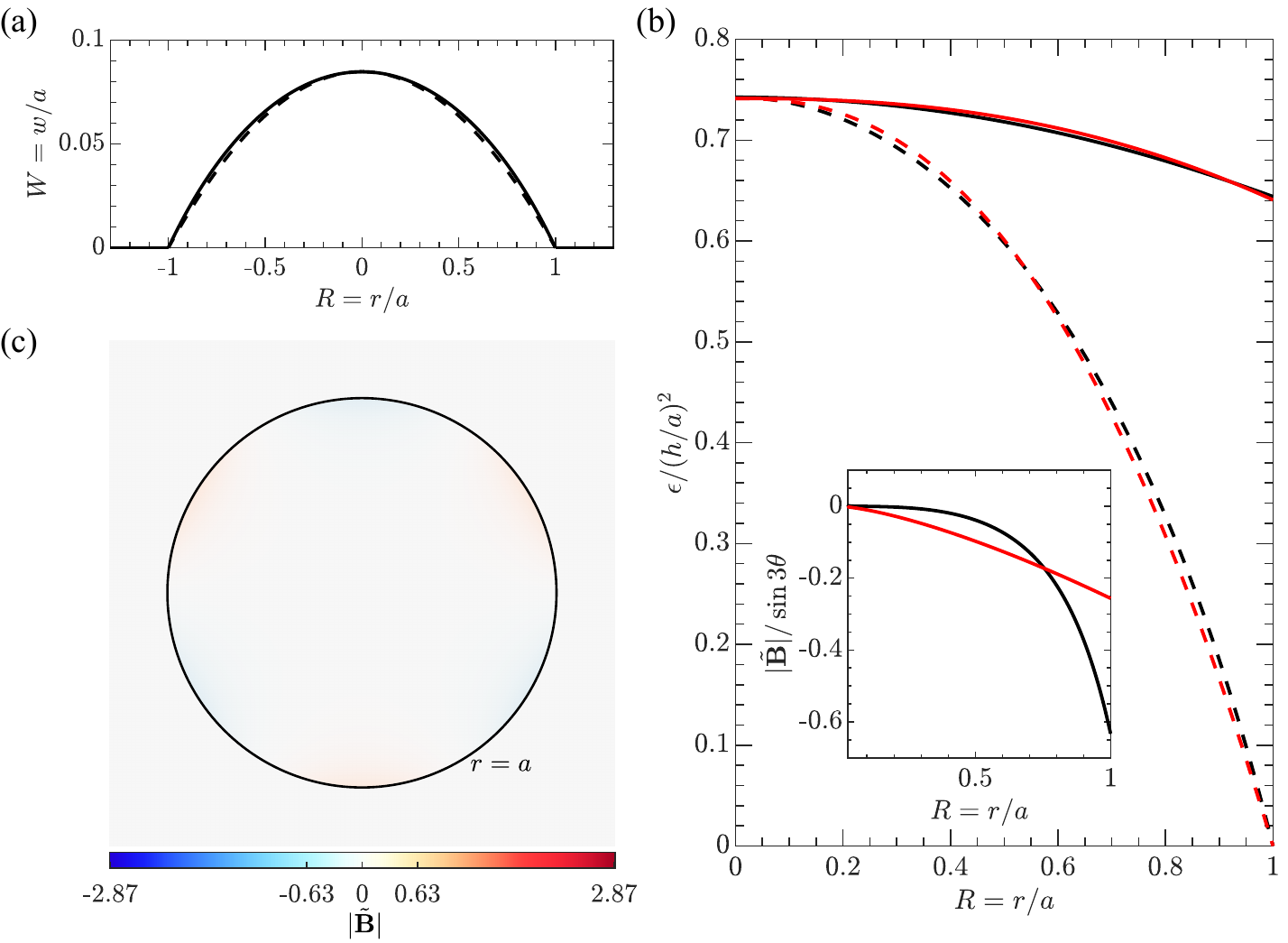}
    \caption{Shape, strain, and pseudomagnetic fields (PMFs) in bubbles with adhesive, no-sliding boundary conditions. (a) The deflection of a typical bubble is calculated using $\gamma=10^{-4}$. As comparison, the dashed curve presents the spherical cap shape. (b) Rescaled radial (solid curves) and hoop (dashed curves) strain distributions for various $\gamma$. The use of $(h/a)^2$ for rescaling collapse these curves. Black curves are from numerical calculations of this work while red curves are based on a simple analysis given in \eqref{eq:StrainDisCaseI} \cite[][]{Dai2018interface,Blundo2021experimental}. The inset shows the strain-gradient-caused $\Bha/\sin3\theta$ according to \eqref{eq:BPMF1} and \eqref{eq:BPMF2}, where again the strains are based on the numerical results of this work (black curve) and the approximate solution \eqref{eq:PMFDisCaseI} (red curve). (c) The rescaled PMFs \eqref{eq:BPMF2} show three-fold symmetric oscillation along a material circle. The magnitude of the rescaled PMFs (indicated by the color bar) increases from the center to the edge of the bubble with the absolute maximum $\approx0.63$. We use a maximum of 2.87 in the color bar throughout the paper for a comparison of different cases (see panel c in Fig.~\ref{fig:SlidingWrinking1}, \ref{fig:SlidingWrinking2}, and to~\ref{fig:SlidingWrinking3}).}
    \label{fig:NoSliding}
\end{figure}

Equations (\ref{eq:Membrane1Rescaled}--\ref{eq:ContactAngleI}) complete the theory for the problem of no-sliding bubbles. We solve this boundary value problem numerically using the built-in solver \texttt{bvp5c} in \textsc{MATLAB}. We show numerical results about the deflection-adhesion relation in Fig.~\ref{fig:gammaLaw} and the shape, strain, and PMFs in Fig.~\ref{fig:NoSliding}.

Consistent with the scaling analysis in \eqref{eq:gammaLaw}, the main conclusion of Fig.~\ref{fig:gammaLaw} is that the aspect ratio of the bubble is proportional to the fourth power of the strength of the adhesion with respect to the stretching stiffness. This power law was also discussed in previous works such as by \cite{Khestanova2016universal}, \cite{Sanchez2018}, \cite{Dai2018interface}, and \cite{Blundo2021experimental}. Here, the prefactor for this power law is determined numerically:
\begin{equation}
    h/a\approx0.84\gamma^{1/4},
    \label{eq:AspectRatioCaseI}
\end{equation}
which is identical to that was numerically fitted in \cite{Blundo2021experimental}, slightly smaller than $[24(1-\nu)/(35-5\nu)]^{1/4}\approx0.88$ with $\nu=0.165$ in \cite{Dai2018interface,Sanchez2018}, and a bit smaller than $0.97$ as numerically obtained in \cite{Khestanova2016universal} in which sliding is allowed. 

Figure~\ref{fig:NoSliding}a shows the deflection of the thin sheet calculated using $\gamma=10^{-4}$. This deflection shape differs from a spherical cap (solution to Young-Laplace equation) because the membrane tensions or strains are non-uniform in general, as further illustrated in Fig.~\ref{fig:NoSliding}b (black curves: solid for $\epsilon_r$ and dashed for $\epsilon_\theta$). This difference was also observed in \cite{Dai2018interface} that attempted to use $\w(\rt)=h/a(1-\rt^\beta)$ to describe the bubble shape so that the strains can be obtained analytically:
\begin{equation}
\begin{split}
    \epsilon_r&=\frac{\beta(2\beta-1-\nu)}{8(\beta-1)}\frac{h^2}{a^2}\left(1-\frac{1+\nu-2\beta\nu}{2\beta-1-\nu}\rt^{2\beta-2}\right),\\
    \epsilon_\theta&=\frac{\beta(2\beta-1-\nu)}{8(\beta-1)}\frac{h^2}{a^2}\left(1-\rt^{2\beta-2}\right).
    \label{eq:StrainDisCaseI}
\end{split}
\end{equation}
It was found that \eqref{eq:StrainDisCaseI} using $\beta=2$ (a spherical cap shape) could not perfectly match numerical results \cite[see Fig. \ref{fig:NoSliding}a in this work and Fig. 3a in][]{Dai2018interface}. A more recent work by \cite{Blundo2021experimental} has improved the accuracy of \eqref{eq:StrainDisCaseI} by taking $\beta=2.2$ with which we plot the red curves in Fig.~\ref{fig:NoSliding}b. As one may have already realized from the scaling analysis or \eqref{eq:StrainDisCaseI}, an important feature of the strain distribution is that it depends only on the aspect ratio of the bubble. The plots in Fig.~\ref{fig:NoSliding}b have used various $\gamma$ but collapsed after the strains are normalized by the aspect ratio squared. Specifically, for graphene with $\nu=0.165$ the strain at the bubble center is
\begin{equation}
    \epsilon_r(0)=\epsilon_\theta(0)\approx0.74(h/a)^2,
    \label{eq:MaxStrainCaseI}
\end{equation}
which is also the maximum strain the bubble system can achieve.

Despite of an overall good agreement between the approximate analytical solution \eqref{eq:StrainDisCaseI} and numerical results in the sense of strain distribution, we found that \eqref{eq:StrainDisCaseI} is not precise enough to predict the \emph{strain gradient distribution} and hence the PMFs. For example, combining \eqref{eq:StrainDisCaseI} and \eqref{eq:BPMF1} leads to 
\begin{equation}
    \Bha=\tfrac{1}{2}\sin{3\theta}(1+\nu)\beta(2-\beta)\rt^{2\beta-3},
    \label{eq:PMFDisCaseI}
\end{equation}
which gives the rescaled PMFs (red curve) in Fig.~\ref{fig:NoSliding}b. This, however, has appreciable deviation from the numerically determined $\Bha$ (black curve) PMFs. 

We further show the full field PMFs by considering their angular dependence in a circular bubble in Fig.~\ref{fig:NoSliding}c. The three-fold symmetry reported in the literature is observed \cite[][]{Settnes2016,Qi2014} in our modeling result. It is also found that the absolute magnitude of the rescaled PMF increases from the center to the edge of the bubble with a maximum value of 0.63, i.e.
\begin{equation}
    \mathrm{max} |\BPMF| \approx 0.63A_\mathrm{PMF}h^2/a^3.
    \label{eq:MaxBCaseI}
\end{equation} These results, including aspect ratio-adhesion relation, max strain, and PMFs, and so on, are also summarized in Table \ref{tab}. 

We conclude this subsection by revisiting \eqref{eq:PMFDisCaseI}. In particular, \eqref{eq:PMFDisCaseI} suggests that $\beta$ is a rough geometrical indicator of the magnitude of PMFs in a bubble: the more the bubble shape deviates from the spherical cap ($\beta=2$) the stronger the generated PMFs are. We have seen that the deviation is moderate for no-sliding bubbles ($\beta\approx2.2$ though subject to some quantitative error in predicting PMFs). A natural question is how the bubble geometry and hence the PMFs would be modified by the sliding of the sheet since this is very likely to occur for slippery 2D crystals \cite[see reviews by][]{Hod2018,Liechti2019,Dai2020mechanics}. To answer this question is goal of the rest of this paper.

\section{Sliding and wrinkling in the suspended region\label{sec:Wrinkling1}}

\subsection{Regime}
$$\Sea\gg1,\quad\Kh\gg1,\quad\Ks\ll1$$

Following Section 3 (no sliding), we then consider the opposite limit of the shear behavior---a small shear stress that gives rise to a large sliding ability of the sheet ($\Sea\gg1$). In this limit the interface can be treated as frictionless \cite[][]{Dai2021poking}. The sliding is inward, which causes a material circle to shrink radially and thus be compressed circumferentially. The behavior of thin sheets in response to such compression highly depends on their wrinkling ability, as discussed in Sec.~\ref{sec:WrinklingAbility}. In the suspended region, we focus on systems with large FvK numbers $(\Kh\gg1)$ to reasonably neglect bending inside the bubble, which implies that the thin sheet would wrinkle to release $N_\theta^{bare}$ in favor of the trivial $N_\theta^{res}$. In the supported region, however, this section considers a low bending ability ($\Ks\ll1$) so that the sheet remains planar and the compressive stress due to the adhesion and sliding (up to $\sim Yh^2/a^2\sim Y\gamma^{1/2}$) is maintained in the whole supported region (see the schematic illustration in Fig.~\ref{fig:Schematic}c).

\subsection{Theory}

As illustrated in Fig.~\ref{fig:Schematic}c, the problem contains three regions: a tensile core $0<\rt<\rin=\ell_I/a$, a wrinkled zone $\rin<\rt<1$, and the supported region $\rt>1$. In the tensile core, the equilibrium equation \eqref{eq:Membrane1Rescaled} and compatibility condition \eqref{eq:Membrane2Rescaled} are still applicable for solving $\psit$ and $\w$. We then discuss the other two regions.

\emph{The wrinkled zone $\rin<\rt<1$.} In large $\Kh$ systems, the formation of wrinkles largely relaxes the compressive stress since $\Nt^{res}/\Nr\sim\Kh^{-1/2}\ll1$. We then employ the tension field theory in this region \cite[][]{Pipkin1986relaxed,Steigmann1990tension,Vella2018regimes,Davidovitch2021indentation} that assumes $\Nt=0$. The in-plane force balance \eqref{eq:FvKEqns2} becomes ${\dd (\rt\Nr)}/{\dd \rt}=0$, leading to
\begin{equation}
    \Nr={C}/\rt\quad\text{and}\quad\psit=C
    \label{eq:TFTSuspended}
\end{equation}
with $C$ a not-yet-known constant. With this the deflection of the thin sheet can be solved using the vertical force balance \eqref{eq:Membrane1Rescaled}:
\begin{equation}
    \w=\frac{\p}{6C}(1-\rt^3).
    \label{eq:ShapeCaseII}
\end{equation}

\emph{The supported region $\rt>1$.} The vanished shear stress and the planar state of the thin sheet returns to the Lamé problem\cite[][]{Sadd2009}; Its solutions read
\begin{equation}
    \Nr=C/\rt^2\quad\text{and}\quad\Nt=-C/\rt^2,
    \label{eq:Lame}
\end{equation}
where we used $\Nr(1^-)=\Nr(1^+)=C$. In addition, we assumed that the size of the thin sheet is large enough ($\rsheet\gg1$) and neglected any residual stress in the far field. As a result of these assumptions, the problem is largely simplified, which is to solve \eqref{eq:Membrane1Rescaled} and \eqref{eq:Membrane2Rescaled} with three unknowns: the pressure $\p$, the size of the tensile core $L_I$, and a constant $C$. 

\emph{Boundary and matching conditions.} Six conditions are required to complete the theory. Four straightforward ones are the zero displacement at the bubble center and the continuity of the radial stress, hoop stress, and vertical displacement at the edge of the tensile core:
\begin{equation}
   \lim_{\rt\to0}\left(\rt\psit'-\nu \psit\right)=0,\quad\psit(\rin^-)=C,\quad\psit'(\rin^-)=0,\quad\w(\rin^-)=\frac{\p}{6C}\left(1-\rin^3\right).
    \label{eq:BC_CaseII_Rescale}
\end{equation}
The fifth condition is associated with the in-plane displacement across the wrinkled region. In particular, the kinematics \eqref{eq:Kinematics} and Hooke's law \eqref{eq:Hooke} requires 
\begin{equation*}
    \epsilon_r=\frac{\dd \ut}{\dd \rt}+\frac{1}{2}\left(\frac{\dd \w}{\dd \rt}\right)^2=\frac{C}{\rt}
\end{equation*}
for $\rin<\rt<1$, which, together with \eqref{eq:BC_CaseII_Rescale}, give rise to
\begin{equation}
    40C^3\left(\ln{\rin}-1\right)+\p^2\left(1-\rin^5\right)=0.
    \label{eq:BCDisplCaseII}
\end{equation}
Interestingly, Poisson's ratio does not appear in \eqref{eq:BCDisplCaseII}. (Indeed our numerical results shown in the next section for $\rin$ and $h/a-\gamma$ relation are invariant to $\nu$ in this case.) 

Finally, the elastoadhesive interaction brings a slope-jump condition \eqref{eq:ContactAngle} at $\rt=1$, which is found to take the same form as \eqref{eq:ContactAngleI}:
\begin{equation}
    \Nr(1-\cos\vartheta)=\gamma-\tfrac{1}{2}\Nr\epsilon_r
    \label{eq:ContactAngleII}
\end{equation}
but with $\epsilon_r=\epsilon_r(1^-)$. In this case $\epsilon_r(1^-)\neq\epsilon_r(1^+)$ due to the wrinkling that takes place only in the suspended region.

\begin{figure}[!t]
    \centering
    \vspace{0.0cm}
    \includegraphics[width=14.5cm]{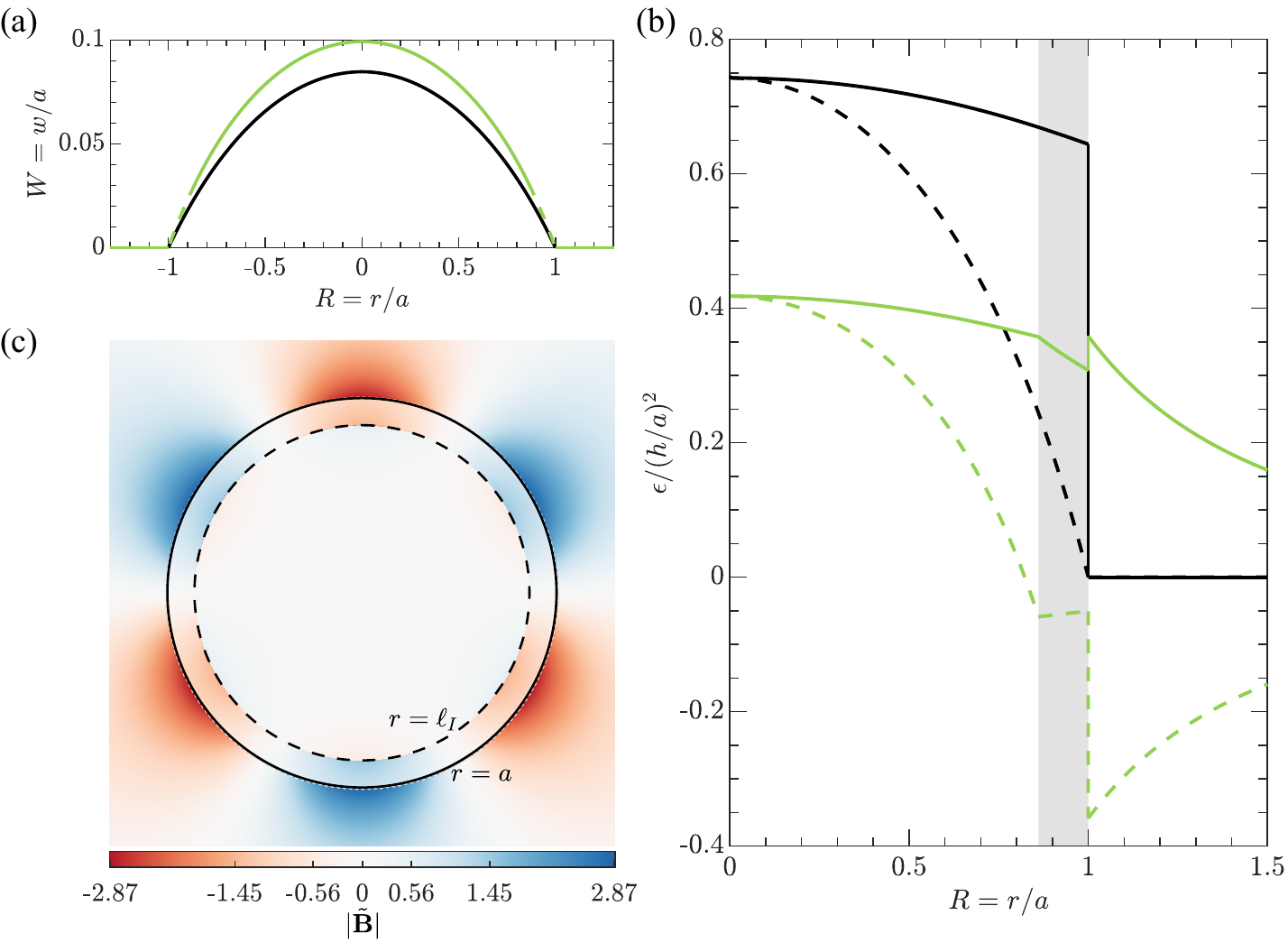}
    \caption{Shape, strain, and pseudomagnetic fields (PMFs) in bubbles with a wrinkled zone $\ell_I<r<1$ due to the adhesive and sliding boundary conditions. Black curves: no-sliding (Sec.~\ref{sec:NoSliding}); Green curves: sliding and wrinkling in the suspended region (Sec.~\ref{sec:Wrinkling1}). (a) The deflection of a typical bubble is calculated using $\gamma=10^{-4}$. The dashed part of the green curve denotes the wrinkled zone. (b) Rescaled radial (solid curves) and hoop (dashed curves) strain distributions for various $\gamma$. The gray-shaded region highlights the wrinkled zone. (c) The rescaled PMFs \eqref{eq:BPMF2} with magnitude encoded by the color. The wrinkled region is the annular region between the dashed and solid circle.}
    \label{fig:SlidingWrinking1}
\end{figure}

\subsection{Numerical results}
We solve the problem \eqref{eq:Membrane1Rescaled} and \eqref{eq:Membrane2Rescaled} with boundary and matching conditions \eqref{eq:BC_CaseII_Rescale}--\eqref{eq:ContactAngleII} numerically and show the numerical results about the deflection-adhesion relation in Fig.~\ref{fig:gammaLaw} and the shape, strain, and PMFs in Fig.~\ref{fig:SlidingWrinking1}.

Figure 2 clearly shows that when sliding and wrinkling occur the power-law \eqref{eq:gammaLaw} still works and exhibits an increased prefactor (compared with the no-sliding case):
\begin{equation}
    h/a\approx0.98\gamma^{1/4}.
    \label{eq:AspectRatioCaseII}
\end{equation}
This can also be seen in Fig.~\ref{fig:SlidingWrinking1}a where under the same $\gamma=10^{-4}$, the sliding-and-wrinkling bubble (green curve) produces larger deflections than the no-sliding bubble (black curve). These observations indicate that the sliding in the supported region and the wrinkling in the suspended region make the sheet less resistant to the vdW adhesive forces. It is worth noting that compared with the wrinkling effect, the sliding effect is much more important: the prefactor is $\sim0.96$ in an imaginary situation in which the sheet slides but somehow the compressive stresses are not relaxed (no wrinkling). The reason for the unimportant role of wrinkling is that \eqref{eq:AspectRatioCaseII} embodies a competition between the adhesion energy and the overall elastic strain energy. But the wrinkling only impacts the stress state (particularly $\Nt$) in a very limited region of the bubble. To provide a quantitative insight into this limited effect, we use the size of the unwrinkled, tensile core:
\begin{equation}
    \ell_I=0.86a,
\end{equation}
which is found invariant to the adhesion strength of the interface and even the Poisson's ratio of the sheet. %The prefactor numerically determined here is slightly different from previous reports that accounted for the sliding effect (but neglected the wrinkling effect), say $0.97$ in a numerical model by \cite{Khestanova2016universal} and $(6/5)^{1/4}\approx1.05$ in the approximation model by \cite{Dai2018interface} and \cite{Sanchez2018}; 

Though not significantly affecting the overall elastic energy, the wrinkle formation can regularize the geometry of the bubble as well as the strain distribution in the wrinkled zone $\rin<\rt<1$ (see the gray-shaded region of green curves in Fig.~\ref{fig:SlidingWrinking1}b) in an effective way. Revisiting \eqref{eq:ShapeCaseII} we find that the out-of-plane deformation of the bubble in the wrinkled region is different from the spherical cap shape qualitatively. Besides, the strain gradient is amplified by wrinkling (comparing the green curves with the no-sliding black curves in the shaded region in Fig.~\ref{fig:SlidingWrinking1}b). As a result, the magnitude of PMFs inside the bubble is found to reach its maximum at the inner boundary of the wrinkled zone ($\rt=\rin^+$, see Fig.~\ref{fig:SlidingWrinking1}c), which is more than twice of the maximum PMFs in no-sliding bubbles (Fig.~\ref{fig:NoSliding}c)

We end this section by discussing the sliding effect on the strain and PMFs in $\rt>1$. The sliding decreases the strain level in the suspended region and, and at the same time, introduces strain fields to the supported region. Perhaps not surprisingly, the newly introduced strain fields decay quickly with $\rt$ (scaling as $\rt^{-2}$), which is very advantageous for generating giant PMFs. In Fig.~\ref{fig:SlidingWrinking1}c, our numerical results show that the rescaled magnitude of PMFs decreases with increasing $\rt$ in the supported region; The maximum (located at $\rt=1^+$)
\begin{equation}
    \mathrm{max} |\BPMF| \approx 2.87A_\mathrm{PMF}h^2/a^3
    \label{eq:MaxBCaseII}
\end{equation} 
is much greater than what can be obtained in the suspended region. This finding might be useful for future experiments since previous experimental measurements have focused on the bubbles themselves \cite[][]{Levy2010,Jia2019programmable} but ignored the supported region where the strain gradient is more considerable. We also summarize the results about the aspect ratio-adhesion relation, the max strain and PMFs, and so on in Table \ref{tab}.

The wrinkling ability of a homogeneous sheet on the substrate scales as ${s^2}/{t^2}$ with $t$ the thickness since $D\sim Et^3 = Yt^2$. For multilayer 2D crystals, this scaling may still be used for estimation \cite[as long as the bending level is not too significant according to][]{Wang2019bending,Han2020,Ma2021method}. Thus $\Ks\ll1$ could be readily achieved in a $n$-layer 2D material by roughly requiring $n^2\gg1$. For monolayer 2D crystals, however, the bending stiffness comes from a different origin \cite[][]{Lu2009elastic,Zhang2011bending,Wei2013bending,Zelisko2017determining}, usually on the order of $1\mathrm{~eV}$ \cite[see more detailed summaries in][]{Androulidakis2018tailoring,Li2021}. A typical estimation using $Y\sim100\mathrm{~Nm}$ and $s\sim1\mathrm{~nm}$ gives $\Ks\sim10^3\gg1$---a high wrinkling ability of sheets in the supported region. We therefore move on to discuss the wrinkling behavior in $\rt>1$.

\section{Sliding and wrinkling in both suspended and supported regions\label{sec:Wrinkling2}}

\subsection{Regime}
$$\Sea\gg1,\quad\Kh\gg1,\quad1\ll\rsheet^2\ll\Ks\ll\Kh$$

We consider a highly bendable thin sheet ($\Kh\gg1$ and $\Ks\gg1$) with a nearly frictionless sheet-substrate interface ($\Sea\gg1$) and a small ``process zone''($\Ks\ll\Kh$). Recalling the discussion in Sec.~\ref{sec:WrinklingAbility}, whether the thin sheet prefers to wrinkle depends also on the level of the bare stress \eqref{eq:N_bare}, which decays over $\rt$. In this section, we consider an extremely high wrinkling ability of the sheet so that the sheet favors the wrinkling state as long as the stress is compressive even with a substrate underneath. This limiting case requires $|\Nt^{res}|\ll|\Nt^{bare}|$, i.e., $\Ks\gg\alpha^{-2}$ in the whole supported region, where $\alpha$ describes the ratio of the actual, decayed stress at a position $\rt>1$ to the typical stress at $\rt\sim1$. 

Since we expect to find the smallest $\alpha$ (or bare stress) at the outer boundary of the sheet, the specific $\rsheet=r_\mathrm{sheet}/a$ and the boundary condition there become important (Fig.~\ref{fig:Schematic}d). For simplicity, we consider a large-size thin sheet and assume that the thin sheet is pinned at its outer physical boundary: $\rsheet\gg1$ and $\ut(\rsheet)=0$. With these one may expect the smallest $\alpha\sim\rsheet^{-2}$ (based on the Lamé Solution) and then require $\Ks\gg\rsheet^4$. However, we shall show that the wrinkling in the supported region slows the decaying rate of the hoop stress from $\rt^{-2}$ down to $\rt^{-1}$. A more consistent requirement of $|\Nt^{res}|\ll|\Nt^{bare}|$ for any $\rt\lesssim\rsheet$ is actually $\Ks\gg\rsheet^2$.

\begin{figure}[!t]
    \centering
    \vspace{0.5cm}
    \includegraphics[width=14.5cm]{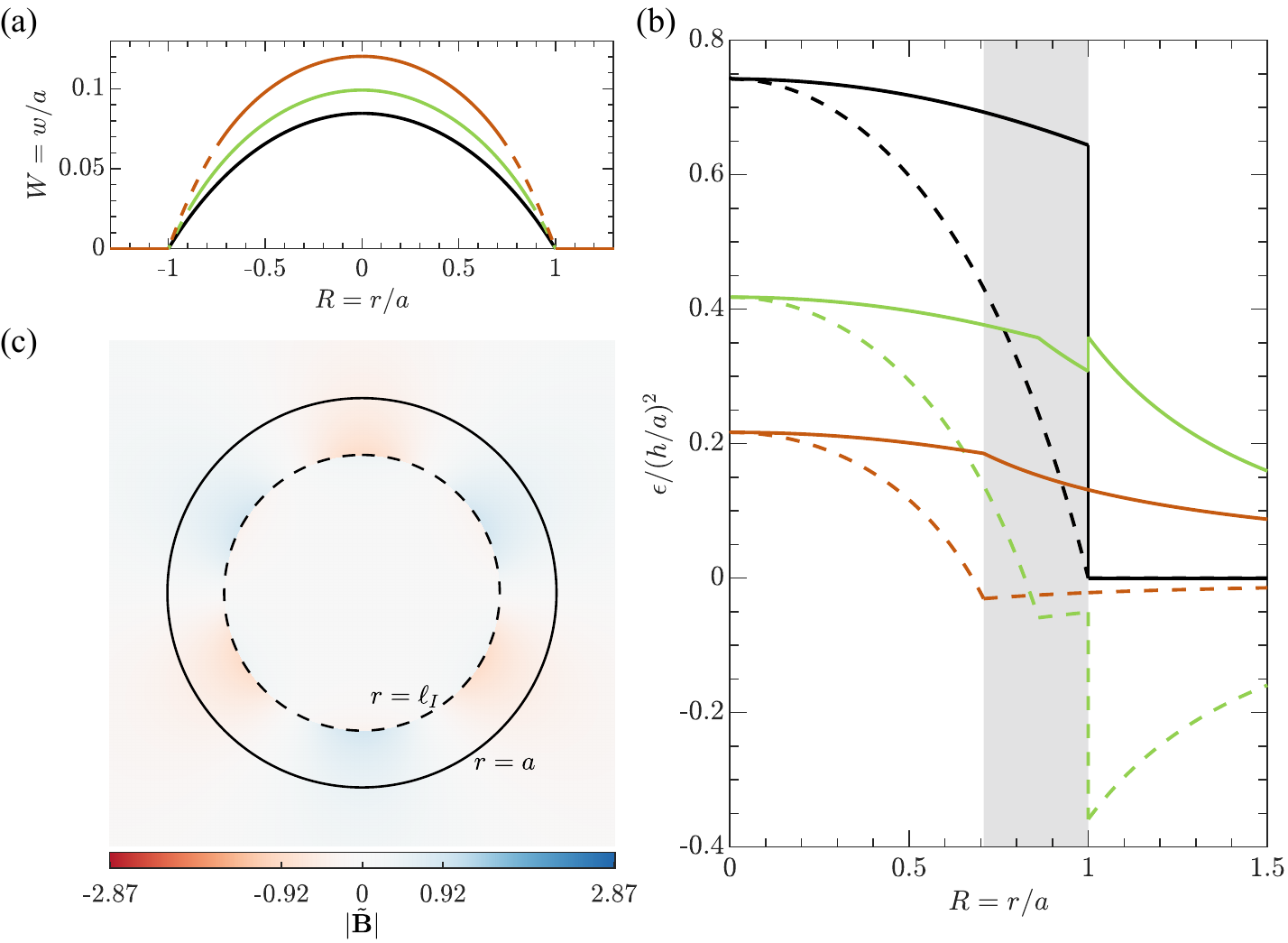}
    \caption{Shape, strain, and pseudomagnetic fields (PMFs) in bubbles that slide and wrinkle in both suspended and supported regions (orange curves). Black curves: no-sliding (Sec.~\ref{sec:NoSliding}). Green curves: sliding and wrinkling only in the suspended region (Sec.~\ref{sec:Wrinkling1}). (a) The deflection of a typical bubble calculated using $\gamma=10^{-4}$. The dashed parts of the curves denote the wrinkled zone. (b) Rescaled, collapsed radial (solid curves) and hoop (dashed curves) strain distributions for various $\gamma$. The gray-shaded region highlights the wrinkled, suspended zone of the orange case. (c) The rescaled PMFs \eqref{eq:BPMF2} with magnitude encoded by the color. The region outside of the dashed circle is in a wrinkled state.}
    \label{fig:SlidingWrinking2}
\end{figure}

\subsection{Theory}

As illustrated in Fig.~\ref{fig:Schematic}d, the problem contains four regions: a tensile core $0<\rt<\rin=\ell_I/a$, a wrinkled, suspended zone $\rin<\rt<1$, a wrinkled, supported zone $1<\rt<\rout=\ell_O/a$, and the outmost unwrinkled, supported region $\rout<\rt<\rsheet$. The mechanics of the two regions in the suspended zone (inside the bubble) should be the same as what we have discussed in Sec.~\ref{sec:Wrinkling2}. So only the two regions outside the bubble are to be discussed.

\emph{The unwrinkled, supported zone $\rout<\rt<\rsheet$.} We fix the displacement at $\rt=\rsheet$, making both radial and hoop stresses tensile in this annulus. We then use the Lamé Solution \cite[][]{Sadd2009}, namely
\begin{equation}
    \Nr=\frac{\ct}{\rt^2}+\frac{1+\nu}{1-\nu}\frac{\ct}{\rsheet^2},\quad\text{and}\quad    \Nt=-\frac{\ct}{\rt^2}+\frac{1+\nu}{1-\nu}\frac{\ct}{\rsheet^2},
    \label{eq:LameCaseIII}
\end{equation}
with the constant $\ct$ to be solved using matching conditions. This form gives the displacement field
\begin{equation}
    \ut=(1+\nu)\ct\left(-\rt^{-1}+\rt{\rsheet^{-2}}\right),
    \label{eq:LameDisplCaseIII}
\end{equation}
satisfying the assumption of pinning: $\ut(\rsheet)=0$. Equation \eqref{eq:LameCaseIII} indicates that the hoop stress is positive at the outer edge of the sheet due to the pinning but eventually becomes negative as $\rt$ moves away from the edge. The location of this boundary is then defined by the location where the hoop stress vanishes, i.e.
\begin{equation}
    \Nt(\rout)=0\quad\implies\quad\rout=\sqrt{\frac{1-\nu}{1+\nu}}\rsheet\quad\text{and}\quad \Nr(\rout)=\frac{2(1+\nu)}{1-\nu}\frac{\ct}{\rsheet^2}.
    \label{eq:Lout}
\end{equation}
Moving further inward the sheet would start to be compressed, leading to wrinkle formation in supported region.

\emph{The wrinkled, supported region $1<\rt<\rout$.} Again, the wrinkling relieves the bare hoop stress and the residual hoop stress $\Nt^{res}$ scales as $C\Ks^{-1/2}$ (see \ref{eq:N_res2}) where $C\sim h^2/a^2$ is the constant Airy stress function in the suspended region \eqref{eq:TFTSuspended}. Immediately, $\Nr/\Nt^{res}\gtrsim \ct\Ks^{1/2}/(C\rsheet^2)\sim(\Ks/\rsheet)^{1/2}\gg1$ where we used $\ct\sim C\rsheet$ which we prove now. Particularly, the negligible residual hoop stress (compared to the tensile radial stress) justifies the application of the tension field theory in this region, leading to solutions similar to \eqref{eq:TFTSuspended}:
\begin{equation}
    \Nt=0\quad\text{and}\quad\Nr=\frac{2(1+\nu)}{1-\nu}\frac{\ct}{\rsheet}\frac{1}{\rt}.
    \label{eq:TFTSupported}
\end{equation}
The coefficient in \eqref{eq:TFTSupported} is determined by the continuity of radial stress at $\rt=\rout$. Similarly, matching the radial stress at $\rt=1$ can give the self-consistent result:
\begin{equation}
    C=\frac{2(1+\nu)}{1-\nu}\frac{\ct}{\rsheet},
\end{equation}i.e., $\ct\sim C\rsheet$. Similar to that in Sec.~\ref{sec:Wrinkling2}, the problem here is to solve \eqref{eq:Membrane1Rescaled} and \eqref{eq:Membrane2Rescaled} as well as the three unknowns: $\p$, $L_I$, and $C$.

\emph{Boundary and matching conditions.} Still applicable are the zero displacement at the bubble center and the continuity of the radial stress, hoop stress, and vertical displacement at the edge of the tensile core provided in \eqref{eq:BC_CaseII_Rescale}. However, the condition of continuous in-plane displacement in \eqref{eq:BCDisplCaseII} should be changed due to the presence of wrinkles in the supported region. Following a similar concept that is utilized to derive \eqref{eq:BCDisplCaseII}, we obtain
\begin{equation*}
    \ut(\rin)-\ut(\rout)=\int_{\rin}^{\rout} \frac{C}{\rt}\dd \rt-\int_{\rin}^1 \frac{1}{2}\left(\frac{\dd \w}{\dd \rt}\right)^2\dd \rt,
    \label{eq:BCDisplsCaseIII_original}
\end{equation*}
which can also be expressed as
\begin{equation}
    40C^3\left(\ln{\rin/\rout}-\nu-\nu\rout/\rsheet\right)+\p^2\left(1-\rin^5\right)=0.
    \label{eq:BCDisplCaseIII}
\end{equation}
Unlike \eqref{eq:BCDisplCaseII} in Sec.~\ref{sec:Wrinkling1}, \eqref{eq:BCDisplCaseIII} depends on the Poisson's ratio of the sheet as well as the physical size of the sheet. Also changed is the slope-jump condition at $\rt=1$, which according to \eqref{eq:ContactAngle} can be expressed as
\begin{equation}
    \Nr(1-\cos\vartheta)=\gamma.
    \label{eq:ContactAngleIII}
\end{equation}
Interestingly, the energy term in \eqref{eq:ContactAngle} cancels out as the sheet wrinkles on both sides of the bubble edge---making \eqref{eq:ContactAngleIII} very similar to the contact angle of a droplet on a substrate (the surface tensions of the drop and substrate are replaced by elastic membrane tensions).

\subsection{Numerical results}

Equation \eqref{eq:BCDisplCaseIII} renders the problem dependent on the actual size of the thin sheet. In numerics, we use $\rsheet=100$ to solve \eqref{eq:Membrane1Rescaled} and \eqref{eq:Membrane2Rescaled} subjected to boundary and matching conditions \eqref{eq:BC_CaseII_Rescale}, \eqref{eq:BCDisplCaseIII} and \eqref{eq:ContactAngleIII}. Nevertheless, as might be expected, the numerical results such as the deflection-adhesion relation, strain and size of the tensile core depend on $\rsheet$ very weakly (logarithmically) when $\rsheet\gg1$. For example, a fitting suggests $\rin\sim O(1)-\log{(\log \rsheet)}$ though $\rout\sim\rsheet$ \eqref{eq:Lout}.

The main results for this case are summarized in Table~\ref{tab}. We find $h/a\sim\gamma^{1/4}$ relation remains durable (also see Fig.~\ref{fig:gammaLaw}). Its prefactor increases slightly (compared with other cases in the preceding sections, also see Fig.~\ref{fig:SlidingWrinking2}a) due to the wrinkling in the supported region. This observation may be expected since wrinkling is a process to relax stresses or reduce the sheet's resistance to the sheet-substrate adhesion. A unique feature of the calculated strain distributions is that there is no jump across the bubble edge at $\rt=1$. This is because the sheet is allowed to slide without friction (continuous $\Nr$) and the wrinkling occurs both inside and outside the bubble (continuous $\Nt$). As shown in Fig.~\ref{fig:SlidingWrinking2}, this feature also gives a continuous distribution of the PMFs in the majority of the sheet (i.e., the wrinkled zone $\rin<\rt<\rout$). The maximum PMFs in this case is found to locate at the inner boundary of the wrinkled zone: $\mathrm{max} |\BPMF| \approx 0.92A_\mathrm{PMF}h^2/a^3$ at $\rt=\rin^+$.

\section{Sliding and wrinkling: the effect of sheet-substrate normal interactions\label{sec:Wrinkling3}}

\begin{figure}[!ht]
    \centering
    \vspace{0.5cm}
    \includegraphics[width=14.5cm]{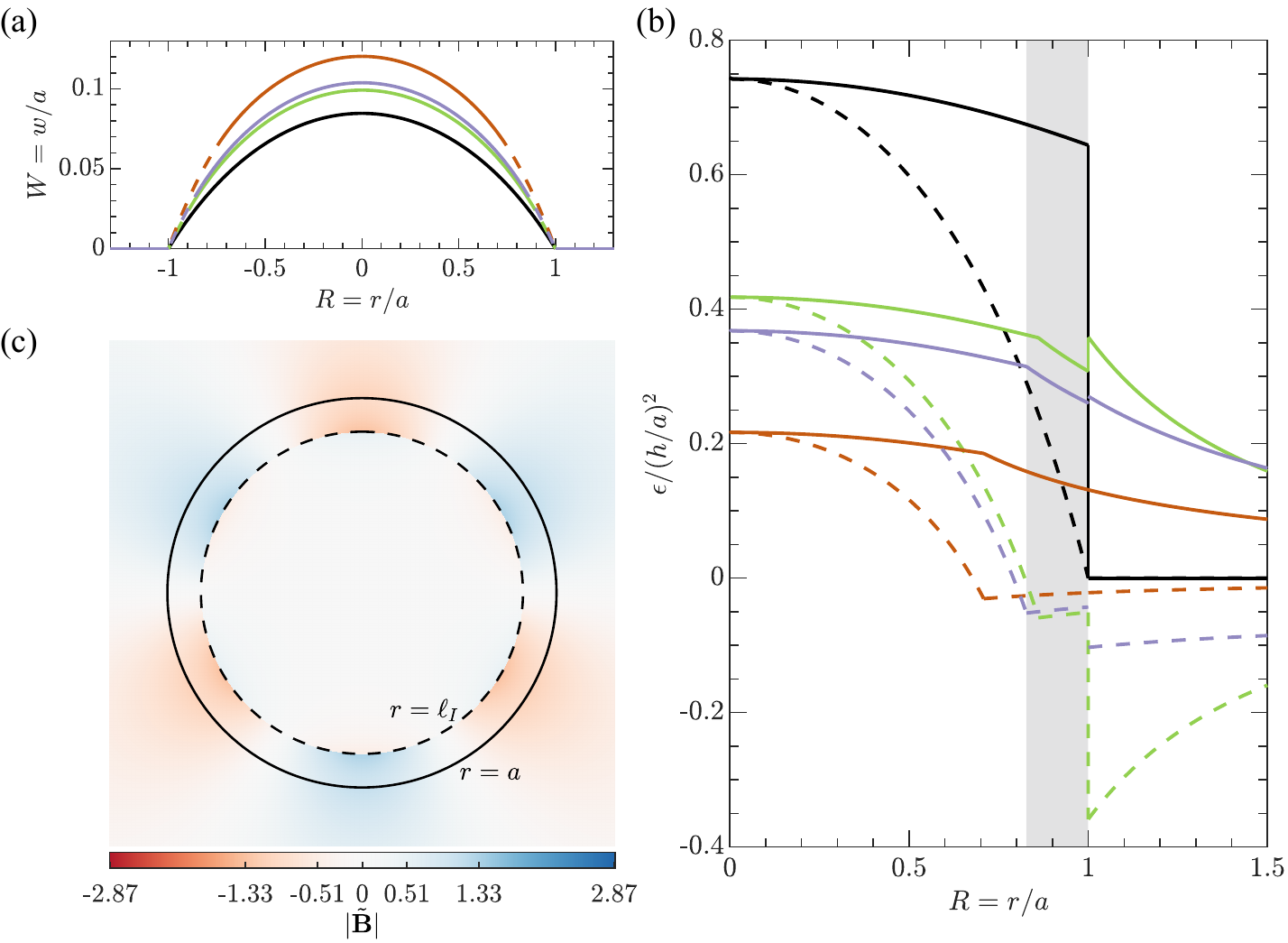}
    \caption{Shape, strain, and pseudomagnetic fields (PMFs) in bubbles that slide and wrinkle in both suspended and supported regions with the consideration of the sheet-substrate adhesion (purple curves, calculated using $\rsheet=100$ and $\Ks=\rsheet^2/10$). Black curves: no-sliding (Sec.~\ref{sec:NoSliding}). Green curves: sliding and wrinkling only in the suspended region (Sec.~\ref{sec:Wrinkling1}). Orange curves: sliding and wrinkling in both suspended and supported regions (with negligible sheet-substrate adhesion) ($\rsheet=100$, Sec.~\ref{sec:Wrinkling2}). (a) The deflection of a typical bubble is calculated using $\gamma=10^{-4}$. The dashed parts of the curves denote the wrinkled zone. (b) Rescaled, collapsed radial (solid curves) and hoop (dashed curves) strain distributions for various $\gamma$. The gray-shaded region highlights the wrinkled zone in the suspended area of the orange case. (c) The rescaled PMFs \eqref{eq:BPMF2} with magnitude encoded by the color. The region outside of the dashed circle is in a wrinkling state.}
    \label{fig:SlidingWrinking3}
\end{figure}

$$\Sea\gg1,\quad\Kh\gg1,\quad1\ll\Ks\lesssim\rsheet^2,\quad \Ks\ll\Kh$$

Finally we discuss the regime of moderately high wrinkling ability of the sheet on its adhesive substrate: $1\ll\Ks\lesssim\rsheet^2$, which lies in between the regimes in Sec.~\ref{sec:Wrinkling1} and Sec.~\ref{sec:Wrinkling2}. Again we enforce $\Ks\ll\Kh$ to ensure a small process zone. Comparing $\Nt^{bare}$ with $\Nt^{res}$ in this regime suggests that the supported region is composed of a region close to the edge of the bubble where $|\Nt^{bare}|>|\Nt^{res}|$ and an outer region where $|\Nt^{bare}|<|\Nt^{res}|$. A key mechanics problem in this regime is then to explicitly consider the $\Nt^{res}$ that was neglected in Sec.~\ref{sec:Wrinkling1} (the regime of low wrinkling ability so no wrinkling in the supported region) and Sec.~\ref{sec:Wrinkling2} (the regime of extremely high wrinkling ability so that $|\Nt^{res}|/|\Nt^{bare}|\to0$ in the wrinkled, supported region). We should note that this particular regime was also discussed in the poking problem of a thin sheet on a substrate with vdW interactions by \cite{Davidovitch2021indentation}.

Similar to Sec.~\ref{sec:Wrinkling2}, the problem contains a tensile core $0<\rt<\rin$, a fully wrinkled zone $\rin<\rt<1$ in the suspended region, and a ``partially'' wrinkled zone $1<\rt<\rout$ plus an unwrinkled zone $\rout<\rt<\rsheet$ in the supported region. The main differences come from the `partially' wrinkled zone where a finite $\Nt^{res}$ should be taken into account. To address this, we follow the model by \cite{Davidovitch2021indentation} that approximates this residual stress using \eqref{eq:N_res} and \eqref{eq:lambdaLaw}:
\begin{equation}
    N_\theta=-2\left(D\Ksup\right)^{1/2}\quad\implies\quad\Nt=-2\left(\gamma/\Ks\right)^{1/2},
    \label{eq:ResidualStress}
\end{equation} for $1<\rt<\rout$. This then adds an equibiaxial compression into typical tension field analysis such as \eqref{eq:TFTSupported}:
\begin{equation}
    \Nr=-2\left(\gamma/\Ks\right)^{1/2}+\frac{\bar C}{\rt},
    \label{eq:ITFT}
\end{equation}
with $\bar C$ an unknown constant. This approximation is equivalent to that used in the inverted tension field theory \cite[][]{Davidovitch2019geometrically} that was designed for thin solids under geometrically incompatible confinement with a traction-free boundary condition.

The problem here is to solve \eqref{eq:Membrane1Rescaled}, \eqref{eq:Membrane2Rescaled} and the unknown $\p$, $L_I$, $\rout$, $C$, $\ct$, and $\bar C$ (requiring 9 boundary and matching conditions). In addition to the 4 conditions in \eqref{eq:BC_CaseII_Rescale}, 3 continuity conditions across this partially wrinkled region can be obtained:
\begin{equation}
    \Nr(1^-)=\Nr(1^+),\quad \Nr(\rout^-)=\Nr(\rout^+),\quad\Nt(\rout^-)=\Nt(\rout^+).
\end{equation}
In addition, the continuity of in-plane displacement can be utilized, 
\begin{equation}
    \ut(\rin)-\ut(\rout)=\int_{\rin}^{\rout} \left(\Nr-\nu\Nt\right)\dd \rt-\int_{\rin}^1 \frac{1}{2}\left(\frac{\dd \w}{\dd \rt}\right)^2\dd \rt,
\end{equation}
which is a more generalized form of \eqref{eq:BCDisplsCaseIII_original}. The last condition to close this problem is given by the contact angle at the edge of the bubble that satisfies (see \ref{eq:ACantact4} for more details):
\begin{equation}
    \Nr\left(1-\cos\vartheta\right)=\gamma\left[1+O\left(\Ks^{-1/2}\right)\right]\approx\gamma,
\end{equation}
This form is asymptotically equivalent to \eqref{eq:ContactAngleIII} since the thin sheet has a moderately high wrinkling ability (particularly in regions not too far away from the edge such as the edge of the bubble). 

The numerical results of this regime are shown in Fig.~\ref{fig:gammaLaw} and Fig.~\ref{fig:SlidingWrinking3} (using purple color) and also summarized in Table~\ref{tab}. It is worth noting that the magnitude of the residual stress reflects the effective strength of the normal sheet-substrate interactions. Therefore, the numerical results are expected to depend on $\rsheet$ as well as the $\gamma/\Ks$ used for the calculation. Interestingly, in Fig.~\ref{fig:gammaLaw}, with $\rsheet$ fixed, we find that the coefficient for the $h/a-\gamma^{1/4}$ relation is a function of $\Ks$ only, so are other quantities such as $\rin$ and $\rout$. In addition, the numerical results are tunable between the no-wrinkling limit (Sec.~\ref{sec:Wrinkling1}) and the fully-wrinkled limit (Sec.~\ref{sec:Wrinkling2}), simply by changing the wrinkling ability $\Ks$ of the sheet on the substrate.

\begin{table}[!htb]
\small
    \caption{A summary of geometrical and controlling parameters for 2D crystal bubbles formed after transferred on ``common'' substrates such as polymer, silicon dioxide and so on. By ``common'' we mean that in these systems the interfacial shear resistance, $\tau$, may be on the order of 1 MPa according to previous measurements \cite[as summarized in][]{Dai2020mechanics}. We then used $\tau\sim1\mathrm{~MPa}$ to estimate the sliding parameter, $\Sl$, in different systems listed below. $\Ks$ and $\Kh$ are estimated based on the geometry of the bubble and mechanical properties of the sheet (see Table \ref{Parameters}). Note that monolayer crystals feature a bending stiffness that is independent of their Young's modulus; For multilayer sheets of Young's modulus $E$ and thickness $t$, however, we used $D\sim Et^3/12=Yt^2/12$ as the error caused by using this expression is rather quantitative in axisymmetric, multilayer systems \cite{Ma2022bending}.}
    \label{Experiments_Regular}
\begin{center}
\begin{tabular}{ lcccccc }
        \midrule
        \midrule
 {Materials} & {Radius $a$} & Height $h$ & $h/a$ & $\Sl$ & $\Ks$ & $\Kh$ \\
        \midrule
\rowcolor{Col2}
$\mathrm{G/PVA}$ \cite[][]{Pandey2022polymer}&$\sim20\mathrm{~\mu m}$&$\sim0.3\mathrm{~\mu m}$&$\sim0.02$&$\sim1$&$\sim10^{-6}$&$\sim0.1$\\
        \rowcolor{Gray}
$\mathrm{G/SiO_2}$ \cite[][]{Koenig2011}&$2.5-4.0\mathrm{~\mu m}$&$0.2-0.6\mathrm{~\mu m}$&$\sim0.1$&$\sim1$&$10^{-2}-10^2$&$10^3-10^8$\\
        \rowcolor{Gray}
$\mathrm{G/SiO_2}$ \cite[][]{Georgiou2011}&$\sim 7\mathrm{~\mu m}$&$\sim0.9\mathrm{~\mu m}$&$\sim0.1$&$\sim1$&$\sim10^{2}$&$\sim10^9$\\
$\mathrm{G/SiO_2}$ \cite[][]{Sanchez2018}&$50-250\mathrm{~nm}$&$2-12\mathrm{~nm}$&$\sim0.05$&$\sim10$&$\sim10^2$&$\sim10^4$\\
\rowcolor{Col1}
$\mathrm{G/SiO_2}$ \cite[][]{Zhang2020construction}&$8-38\mathrm{~nm}$&$0.6-3.6\mathrm{~nm}$&$\sim0.1$&$\sim10^2$&$\sim10^2$&$10^2-10^4$\\
\rowcolor{Col2}
$\mathrm{G/SiO_2}$ \cite[][]{Wang2022}&$0.1-1\mathrm{~\mu m}$&$\lesssim10\mathrm{~nm}$&$\lesssim0.05$&$\lesssim1$&$\ll1$&$\lesssim0.1$\\
\rowcolor{Gray}
$\mathrm{MoS_2/SiO}_x$ \cite[][]{Lloyd2017adhesion}&$5-8\mathrm{~\mu m}$&$0.2-1.2\mathrm{~\mu m}$&$<0.2$&$\sim1$&$10^{-2}-10$&$10^5-10^9$\\
$\mathrm{MoS_2/SiO_2}$ \cite[][]{Sanchez2018}&$50-150\mathrm{~nm}$&$3-6\mathrm{~nm}$&$\sim0.05$&$\sim10$&$\sim10$&$\sim10^3$\\
\rowcolor{Gray}
$\mathrm{MoS_2/SiO_2}$ \cite[][]{Luo2020}&$\sim10\mathrm{~\mu m}$&$\sim1\mathrm{~\mu m}$&$\sim0.1$&$\sim1$&$\sim10$&$\sim10^3$\\
$\mathrm{MoS_2/SiO_2}$ \cite[][]{Di2020nanoscale}&$\sim437\mathrm{~nm}$&$\sim69\mathrm{~nm}$&$\sim0.16$&$\sim10$&$\sim10$&$\sim10^5$\\
\rowcolor{Col2}
$\mathrm{MoS_2/SiO_2}$ \cite[][]{Wang2022}&$\sim100\mathrm{~nm}$&$\lesssim10\mathrm{~nm}$&$\lesssim0.07$&$\lesssim10^2$&$\ll1$&$\lesssim1$\\
$\mathrm{MoS_2/Al_2O_3}$ \cite[][]{Sanchez2018}&$25-60\mathrm{~nm}$&$2-5\mathrm{~nm}$&$\sim0.08$&$\sim10$&$\sim10$&$\sim10^3$\\
$\mathrm{WS_2/Gold}$ \cite[][]{Darlington2020facile}&$\sim70\mathrm{~nm}$&$\sim14\mathrm{~nm}$&$\sim0.2$&$\sim10^2$&$\sim10$&$\sim10^4$\\
        \midrule
        \midrule
\end{tabular}
\end{center}
\end{table} 

\begin{table}[!htb]
\small
    \caption{A summary of geometrical and mechanical parameters for 2D crystal bubbles formed by introducing/trapping small molecules between the crystal and the substrate on which it is grown (no transfer process involved). The interfacial shear resistance in this circumstance is not clear so the sliding parameter is not provided. However, bubbles created in this way are often of small sizes --- their behavior may be better understood via more sophisticated interfacial models (the assumption of a constant shear resistance breaks down anyway).}
    \label{Experiments_Growth}
\begin{center}
\begin{tabular}{ lccccc }
        \midrule
        \midrule
 {Materials} & {Radius $a$} & Height $h$ & $h/a$ & $\Ks$ & $\Kh$ \\
        \midrule
$\mathrm{G/Cu}$ \cite[][]{Aslyamov2022universal}&$20-100\mathrm{~nm}$&$5-30\mathrm{~nm}$&$0.1-0.4$&$\sim10^2$&$10^4-10^6$\\
$\mathrm{G/Ge}$ \cite[][]{Jia2019programmable}&$20-140\mathrm{~nm}$&$3-18\mathrm{~nm}$&$\sim0.08$&$\sim10^2$&$10^4-10^5$\\ 
\rowcolor{Col1} 
$\mathrm{G/Pt}$ \cite[][]{Levy2010}&$\sim2\mathrm{~nm}$&$\sim0.4\mathrm{~nm}$&$\sim0.2$&$\sim10^2$&$\sim10^2$\\
\rowcolor{Col1}       
$\mathrm{G/Pt}$ \cite[][]{Villarreal2021breakdown}&$0.5-3.5\mathrm{~nm}$&$0.2-0.6\mathrm{~nm}$&$0.1-1$&$\sim10^2$&$\sim10^2$\\ 
\rowcolor{Col1} 
$\mathrm{G/Ru}$ \cite[][]{Lu2012transforming}&$\sim6\mathrm{~nm}$&$\sim0.5\mathrm{~nm}$&$\sim0.08$&$\sim10^2$&$\sim10^2$\\
        \midrule
        \midrule
\end{tabular}
\end{center}
\end{table}

\begin{table}[!]
\small
    \caption{A summary of geometrical and mechanical parameters for 2D crystal bubbles formed on substrates with atomically smooth surfaces. The interfacial shear resistance in this circumstance may be considered to vanish and hence the sliding parameter $\Sl$ goes infinite.}
    \label{Experiments_Lubricated}
\begin{center}
\begin{tabular}{ lccccc }
        \midrule
        \midrule
 {Materials} & {Radius $a$} & Height $h$ & $h/a$ & $\Ks$ & $\Kh$ \\
        \midrule
$\mathrm{G/CaF_2}$ \cite[][]{Temmen2014hydration}&$\sim100\mathrm{~nm}$&$\sim1\mathrm{~nm}$&$\sim0.01$&$\sim10^2$&$\sim10^3$\\
$\mathrm{G/Diamond}$ \cite[][]{Xuan2013hydrothermal}&$5-30\mathrm{~nm}$&$1-5\mathrm{~nm}$&$\sim0.2$&$\sim10^{2}$&$10^3-10^4$\\
$\mathrm{G/G}$ \cite[][]{Ghorbanfekr2017dependence}&$\sim140\mathrm{~nm}$&$\sim15\mathrm{~nm}$&$\sim0.11$&$\sim10^{2}$&$\sim10^5$\\
$\mathrm{G/G}$ \cite[][]{Hou2021}&$20-140\mathrm{~nm}$&$2-20\mathrm{~nm}$&$\sim0.12$&$\sim10^{2}$&$10^3-10^5$\\
$\mathrm{G/Graphite}$ \cite[][]{Cao2011microscopic}&$\sim100\mathrm{~nm}$&$\sim10\mathrm{~nm}$&$\sim0.1$&$\sim10^{2}$&$\sim10^5$\\
\rowcolor{Col2}
$\mathrm{G/Graphite}$ \cite[][]{An2017graphene}&$\sim100\mathrm{~nm}$&$\sim10\mathrm{~nm}$&$\lesssim0.12$&$\lesssim10^{-2}$&$\lesssim1$\\
$\mathrm{G/hBN}$ \cite[][]{Uwanno2015fully}&$\sim100\mathrm{~nm}$&$\sim10\mathrm{~nm}$&$\sim0.11$&$\sim10^2$&$10^5$\\
$\mathrm{G/hBN}$ \cite[][]{Khestanova2016universal}&$10-400\mathrm{~nm}$&$1-50\mathrm{~nm}$&$\sim0.11$&$\sim10^2$&$10^3-10^5$\\
$\mathrm{G/hBN}$ \cite[][]{Fei2016ultraconfined}&$\sim125\mathrm{~nm}$&$<20\mathrm{~nm}$&$<0.16$&$\sim10^{2}$&$\sim10^5$\\
\rowcolor{Col2}
$\mathrm{G/hBN}$ \cite[][]{Pizzocchero2016hot}&$\sim2\mathrm{~\mu m}$&$\sim120\mathrm{~nm}$&$\sim0.06$&$\sim10^{-5}$&$\sim0.1$\\
\rowcolor{Col2}
$\mathrm{G/hBN}$ \cite[][]{Wang2022}&$10-10^3\mathrm{~nm}$&$\lesssim10\mathrm{~nm}$&$\lesssim0.1$&$\ll1$&$\lesssim1$\\
$\mathrm{G/Ice}$ \cite[][]{Bampoulis2016hydrophobic}&$18-300\mathrm{~nm}$&$3-18\mathrm{~nm}$&$\sim0.06$&$\sim10^2$&$10^4-10^5$\\
$\mathrm{G/MoS_2}$ \cite[][]{Bampoulis2016hydrophobic}&$17-230\mathrm{~nm}$&$3-30\mathrm{~nm}$&$\sim0.13$&$\sim10^2$&$10^4-10^6$\\
$\mathrm{G/ReS_2}$ \cite[][]{Boddison2019flattening}&$\sim100\mathrm{~nm}$&$\sim10\mathrm{~nm}$&$\sim0.06$&$\sim10^2$&$\sim10^5$\\
\rowcolor{Col2}
$\mathrm{G/Sapphire}$ \cite[][]{Wang2022}&$\sim100\mathrm{~nm}$&$\sim10\mathrm{~nm}$&$\lesssim0.05$&$\ll1$&$\lesssim0.1$\\
$\mathrm{hBN/G}$ \cite[][]{Wang2021visualizing}&$\sim40\mathrm{~nm}$&$\sim10\mathrm{~nm}$&$\sim0.25$&$0.1-10$&$10-10^5$\\ 
$\mathrm{hBN/hBN}$ \cite[][]{Khestanova2016universal}&$10-100\mathrm{~nm}$&$1-10\mathrm{~nm}$&$\sim0.11$&$\sim10^2$&$10^3-10^5$\\
\rowcolor{Col2}
$\mathrm{hBN/hBN}$ \cite[][]{He2019isolating}&$0.1-5\mathrm{~\mu m}$&$2-200\mathrm{~nm}$&$\sim0.10$&$\ll1$&$\ll1$\\
$\mathrm{hBN/hBN}$ \cite[][]{Ares2020piezoelectricity}&$\sim130\mathrm{~nm}$&$\sim16\mathrm{~nm}$&$\sim0.12$&$\sim10^2$&$\sim10^5$\\
$\mathrm{hBN/hBN}$ \cite[][]{Blundo2021experimental}&$0.06-4\mathrm{~\mu m}$&$10-10^3\mathrm{~nm}$&$\sim0.11$&$\sim10^2$&$10^5-10^9$\\ 
$\mathrm{hBN/hBN}$ \cite[][]{Ares2021}&$\sim100\mathrm{~nm}$&$\sim10\mathrm{~nm}$&$0.1-0.2$&$\sim10^2$&$\sim10^5$\\
$\mathrm{MoS_2/G}$ \cite[][]{Tyurnina2019strained}&$\sim1\mathrm{~\mu m}$&$\sim100\mathrm{~nm}$&$\sim0.15$&$\sim10$&$\sim10^6$\\
$\mathrm{MoS_2/G}$ \cite[][]{Wang2021visualizing}&$\sim200\mathrm{~nm}$&$\sim30\mathrm{~nm}$&$\sim0.15$&$\sim10$&$\sim10^4$\\ 
$\mathrm{MoS_2/Graphite}$ \cite[][]{Xu2022nanoscale}&$\sim50\mathrm{~nm}$&$\sim7\mathrm{~nm}$&$\sim0.14$&$\sim10^2$&$\sim10^4$\\ 
$\mathrm{MoS_2/hBN}$ \cite[][]{Khestanova2016universal}&$20-200\mathrm{~nm}$&$3-30\mathrm{~nm}$&$\sim0.14$&$\sim10$&$10^3-10^5$\\
$\mathrm{MoS_2/hBN}$ \cite[][]{Tyurnina2019strained}&$\sim1\mathrm{~\mu m}$&$\sim100\mathrm{~nm}$&$\sim0.15$&$\sim10$&$\sim10^6$\\
$\mathrm{MoS_2/hBN}$ \cite[][]{Boddison2019flattening}&$\sim100\mathrm{~nm}$&$\sim10\mathrm{~nm}$&$\sim0.1$&$\sim10$&$\sim10^4$\\
$\mathrm{MoS_2/hBN}$ \cite[][]{Blundo2021experimental}&$\sim100\mathrm{~n m}$&$\sim10\mathrm{~nm}$&$\sim0.12$&$\sim10$&$\sim10^6$\\ 
$\mathrm{MoS_2/MoS_2}$ \cite[][]{Khestanova2016universal}&$0.02-1\mathrm{~\mu m}$&$3-200\mathrm{~nm}$&$\sim0.17$&$\sim10$&$10^3-10^6$\\
$\mathrm{MoS_2/MoS_2}$ \cite[][]{Tyurnina2019strained}&$\sim1\mathrm{~\mu m}$&$\sim100\mathrm{~nm}$&$\sim0.15$&$\sim10$&$\sim10^6$\\
$\mathrm{MoS_2/MoS_2}$ \cite[][]{Tedeschi2019controlled}&$0.1-3\mathrm{~\mu m}$&$10-10^3\mathrm{~nm}$&$\sim0.16$&$\sim10$&$10^4-10^8$\\ 
$\mathrm{MoS_2/MoS_2}$ \cite[][]{Blundo2020engineered}&$\sim1\mathrm{~\mu m}$&$\sim100\mathrm{~nm}$&$\lesssim0.22$&$\sim10$&$\sim10^6$\\ 
$\mathrm{MoS_2/MoS_2}$ \cite[][]{Tan2020direct}&$\sim1\mathrm{~\mu m}$&$\sim100\mathrm{~nm}$&$\sim0.1$&$\lesssim0.1$&$\lesssim10^3$\\
$\mathrm{MoS_2/MoTe_2}$ \cite[][]{Blundo2021experimental}&$\sim1\mathrm{~\mu m}$&$\sim100\mathrm{~nm}$&$\sim0.12$&$\sim10^2$&$\sim10^7$\\ 
$\mathrm{MoS_2/PtSe_2}$ \cite[][]{Tyurnina2019strained}&$\sim1\mathrm{~\mu m}$&$\sim100\mathrm{~nm}$&$\sim0.11$&$\sim10$&$\sim10^6$\\
$\mathrm{MoS_2/WS_2}$ \cite[][]{Tyurnina2019strained}&$\sim1\mathrm{~\mu m}$&$\sim100\mathrm{~nm}$&$\sim0.14$&$\sim10$&$\sim10^6$\\
$\mathrm{MoS_2/WS_2}$ \cite[][]{Blundo2021experimental}&$\sim100\mathrm{~n m}$&$\sim10\mathrm{~nm}$&$\sim0.15$&$\sim10$&$\sim10^6$\\ 
$\mathrm{MoSe_2/MoSe_2}$ \cite[][]{Tedeschi2019controlled}&$0.1-3\mathrm{~\mu m}$&$10-10^3\mathrm{~nm}$&$\sim0.18$&$\sim10$&$10^4-10^8$\\ 
$\mathrm{MoSe_2/MoSe_2}$ \cite[][]{Blundo2021experimental}&$0.06-1\mathrm{~\mu m}$&$10-10^2\mathrm{~nm}$&$\sim0.19$&$\sim10$&$10^4-10^6$\\ 
$\mathrm{MoTe_2/MoTe_2}$ \cite[][]{Tedeschi2019controlled}&$0.1-3\mathrm{~\mu m}$&$10-10^3\mathrm{~nm}$&$\sim0.17$&$\sim10$&$10^4-10^8$\\ 
$\mathrm{WS_2/WS_2}$ \cite[][]{Tedeschi2019controlled}&$0.1-3\mathrm{~\mu m}$&$10-10^3\mathrm{~nm}$&$\sim0.16$&$\sim10$&$10^4-10^8$\\
$\mathrm{WS_2/WS_2}$ \cite[][]{Blundo2021experimental}&$0.2-6\mathrm{~\mu m}$&$10-10^3\mathrm{~nm}$&$\sim0.17$&$\sim10$&$10^4-10^8$\\
$\mathrm{WSe_2/hBN}$ \cite[][]{Shepard2017nanobubble}&$\sim0.5\mathrm{~\mu m}$&$\sim10\mathrm{~nm}$&$\sim0.02$&$\sim10$&$\sim10^4$\\  
$\mathrm{WSe_2/hBN}$ \cite[][]{Darlington2020imaging}&$\sim50\mathrm{~nm}$&$\sim10\mathrm{~nm}$&$\sim0.2$&$\sim10$&$\sim10^4$\\  
$\mathrm{WSe_2/hBN}$ \cite[][]{Blundo2021experimental}&$\sim100\mathrm{~n m}$&$\sim10\mathrm{~nm}$&$\sim0.11$&$\sim10$&$\sim10^6$\\ 
$\mathrm{WSe_2/WS_2}$ \cite[][]{Blundo2021experimental}&$\sim100\mathrm{~n m}$&$\sim10\mathrm{~nm}$&$\sim0.13$&$\sim10$&$\sim10^6$\\
$\mathrm{WSe_2/WSe_2}$ \cite[][]{Tedeschi2019controlled}&$0.1-3\mathrm{~\mu m}$&$10-10^3\mathrm{~nm}$&$\sim0.15$&$\sim10$&$10^4-10^8$\\
$\mathrm{WTe_2/WTe_2}$ \cite[][]{Tedeschi2019controlled}&$0.1-3\mathrm{~\mu m}$&$10-10^3\mathrm{~nm}$&$\sim0.13$&$\sim10$&$10^4-10^8$\\ 
        \midrule
        \midrule
\end{tabular}
\end{center}
\end{table} 

\section{Existing experimental observations}
Having investigated the problem of 2D crystal bubbles in several specific $\{\Sl,\Ks,\Kh\}$ regimes, we summarize existing experimental observations on the shape of these bubbles involving a variety of 2D crystals and substrates in Table \ref{Experiments_Regular}--\ref{Experiments_Lubricated}. In addition, the associated $\{\Sl,\Ks,\Kh\}$ for each set of 2D crystal and substrate are estimated based on the geometry of the bubble and properties of the crystal and the crystal-substrate interface. These parameters directly suggest which specific model presented in this work should be used to analyze the characteristic aspect ratio in different experiments. However, there are also a number of existing experiments (color-marked in these tables) showing parameter regimes that are out of the consideration of this work and thus warrant further studies. We specify such particular experiments as follows:
\begin{itemize}
    \item Gray-colored rows in Table \ref{Experiments_Regular}. The sliding parameter in these experiments is neither in the ``no-sliding'' limit ($\Sl\ll1$) nor in the ``sliding'' limit ($\Sl\gg1$) that are as discussed in this work. The consideration of a finite interfacial shear resistance is required to interpret these experiments though we expect this consideration only causes slight quantitative changes to the limiting cases studied here. 
    \item Cyan-colored rows in Table $\ref{Experiments_Regular}$ and $\ref{Experiments_Growth}$. Bubbles in these experiments can be less than 1 nm in height; $\Kh/\Ks$ as a result is not large enough to validate the JKR-type analysis used in this work. Instead, Maugis-Dugdale-type modeling is more appropriately positioned to describe the mechanics of 2D crystal bubbles when $\Kh$ is on the same numerical order of $\Ks$.
    \item Yellow-colored rows in Table $\ref{Experiments_Regular}$ and $\ref{Experiments_Lubricated}$. A common feature of these experiments is the presence of multilayer 2D crystals. The bending stiffness of the sheet increases dramatically with the increasing number of layers so that a large FvK number ($\Kh$) in the suspended region is not guaranteed in these experiments. The bending effect (neglected in this work) comes into play when $\Kh\lesssim1$, leading to a changing aspect ratio of the bubble with the system size (a characteristic length instead arises). 
\end{itemize}

\section{Conclusions}
In this work, we have studied the deformation of no-slip/slippery thin sheets on adhesive substrates subjected to uniform pressure. We have used a contact angle to describe the jump of the slope at the edge of the bubble as a result of the energetic competition between adhesion and elasticity. One sliding parameter and two wrinkling parameters (one for the suspended region and the other for the supported region) have been found to control the mechanics of the thin sheet. These parameters could be readily estimated using typical elastic properties of the thin sheet and the van der Waals interactions between the sheet and its underlying substrate. We discussed the rich deformation behaviors of thin sheets in several different (limiting) parameter regimes with the consideration of the elastoadhesive interactions. We also showed that sliding is very important in regularizing the strain distribution in the thin sheet and both wrinkling and sliding are important in controlling the PMFs---an electromechanical property of graphene that is sensitive to the in-plane strain gradient. 

Though demonstrated in the specific bubble/pressurization system, our results have established a generic routine for the study of geometry-based strain engineering of 2D crystals in spontaneous systems. Essentially, we predict the equilibrium geometry and the physics tuned by such geometry through solving a boundary value problem with the aid of elastoadhesive boundary conditions. We expect a number of other systems such as bumps, tents, folds, scrolls to be explored by this routine, which may be useful for the deterministic strain engineering of 2D crystals. We also expect rich mechanical behaviors to be revealed in these apparently simple configurations due to the complex interplay among elastoadhesion, sliding and instabilities.

\section*{Declaration of Competing Interest}

The authors declare that they have no known competing financial interests or personal relationships that could have appeared to influence
the work reported in this paper.

\section*{Acknowledgements}

This research was supported by the European Union’s Horizon 2020 research and innovation programme under the Marie Skłodowska-Curie grant agreement No 886028 (Z.D.). Y.R. thanks the Warren A. and Alice L. Meyer Endowed Scholarship in Engineering and N.L. acknowledges support from the Cockrell School of Engineering, both from UT-Austin. We are grateful to Liu Wang, Shuze Zhu, and Daniel Sanchez for the discussion of PMFs.

\appendix
\section{The contact angle\label{Sec:Appendix}}
Let $\E$ be the elastic strain energy density in the thin sheet. We then rewrite \eqref{eq:PotentialEnergy} as
\begin{equation}
    \Pi=\int_0^a\E r\dd r + \int_a^\infty \E r\dd r -p\int_0^a wr\dd r+\tfrac{1}{2}a^2\Gamma,
\end{equation}
where a coeficient of $2\pi$ has been dropped and the selection of $r=\infty$ for simplicity would not change the condition for the contact line. Performing regular variation with $\delta a\neq0$ we have
\begin{equation}
    \delta\Pi=\int_0^a\delta\E r\dd r + \int_a^\infty \delta\E r\dd r -p\int_0^a \delta wr\dd r+a\Gamma\delta a + \E(a^-)a\delta a-\E(a^+)a\delta a.
    \label{eq:DPi}
\end{equation}

We illustrate the steps to derive the ``contact angle'' using the simplest no-sliding case. In this case, the terms outside the bubble in \eqref{eq:DPi} disappear. The elastic strain energy density takes
\begin{equation}
    \E=\tfrac{1}{2}N_r\epsilon_r+\tfrac{1}{2}N_\theta\epsilon_\theta,
\end{equation}
and its variation reads
\begin{equation}
%\begin{split}
    \delta\E = N_r \delta \epsilon_r + N_\theta \delta \epsilon_\theta
    =N_r\delta u' + N_r w'\delta w'+N_\theta\delta u/r. 
    \label{eq:DE}
%\end{split}
\end{equation}
Plugging \eqref{eq:DE} into \eqref{eq:DPi} we obtain 
\begin{gather}
    \delta\Pi=\int_0^a\left[N_\theta-\frac{\dd (r N_r)}{\dd r}\right] \delta u \dd r - \int_0^a\left[\frac{\dd(r N_r w')}{\dd r}+pr\right]\delta w \dd r \\
    + r N_r \delta u|_a + r N_r w'\delta w|_a + a\Gamma\delta a + \E(a)a\delta a.
\end{gather}
$\delta\Pi=0$ can give two equilibrium equations that are identical to \eqref{eq:FvKEqns1} with $D=0$ and \eqref{eq:FvKEqns2}. In addition, grouping the boundary terms leads to
\begin{equation}
    \Gamma-N_r+N_r\left(1-\tfrac{1}{2}{w'}^2\right)-\tfrac{1}{2}N_r\epsilon_r=0,
    \label{eq:ACantact1}
\end{equation}
at $r=a$. Note that $\delta u|_a= \delta u(a) - u'(a)\delta a$ according to the chain rule. In this work we use
\begin{equation}
    \cos\vartheta=1-\tfrac{1}{2}{w'}^2
\end{equation}
for notation so that \eqref{eq:ACantact1} is equivalent to \eqref{eq:ContactAngleI}.

For sliding and wrinkling in the suspended region, the elastic strain energy density $\E=N_r\epsilon_r/2$ in the wrinkled region $\ell_I<r<a$ and $\E=N_r\epsilon_r/2+N_\theta\epsilon_\theta/2$ in the unwrinkled regions. Following the same method used for the no-sliding case we obtain
\begin{equation}
    r N_r \delta u|_{a^-} + r N_r w'\delta w|_{a^-} - r N_r \delta u|_{a^+} + a\Gamma\delta a + \E(a^-)a\delta a-\E(a^+)a\delta a=0.
    \label{eq:BT2}
\end{equation}
Here we did not consider $\delta \ell_I\neq0$ because it has been known that $\partial \Pi /\partial \ell_I=0$ is an inflection point and $\delta\ell_I\neq0$ would bring the condition $N_\theta(\ell_I)=0$ \cite[see more details in][]{Davidovitch2011prototypical,King2012}. The boundary condition \eqref{eq:BT2} can be simplified as 
\begin{equation}
    \Gamma-N_r(a)+N_r(a)\left(1-\tfrac{1}{2}{w'}^2\right)-\tfrac{1}{2}N_r(a)\epsilon_r(a^-)=0,
    \label{eq:ACantact2}
\end{equation}
where we have used $N_r(a^-)=N_r(a^+)$ and $N_r\epsilon_r|_{a^+}=N_\theta\epsilon_\theta|_{a^+}$ (see \ref{eq:Lame}). 

Similarly, for sliding and wrinkling in both suspended and supported regions we have $\E=N_r\epsilon_r/2$ in the wrinkled zone $\ell_I<r<\ell_O$. Neglecting the variation of $\ell_O$ we still obtain \eqref{eq:BT2}, which now is simplified as
\begin{equation}
    \Gamma-N_r(a)+N_r(a)\left(1-\tfrac{1}{2}{w'}^2\right)=0
    \label{eq:ACantact3}
\end{equation}
because of the continuity of both hoop and radial stress at the bubble edge.

For sliding and wrinkling with the consideration of the residual stress, we use a slightly different form of the elastic strain energy density $\E=N_r\epsilon_r/2+N_\theta u/r$ for the wrinkled, supported region: the dropped $1/2$ in the second term is because the residual stress $N_\theta$ is constant \eqref{eq:ResidualStress}, which reflects the material's bending property and the adhesion property (Winkler foundation). The use of this form can reproduce the equilibrium equation \eqref{eq:ITFT}. The boundary terms is also slightly different from other cases: 
\begin{equation}
    \Gamma-N_r(a)+N_r(a)\left(1-\tfrac{1}{2}{w'}^2\right)-\bar U=0,
    \label{eq:ACantact4}
\end{equation}
where $\bar U = N_\theta(a^+) u(a^+)/a- N_r(a)[\epsilon_r(a^+)-\epsilon_r(a^-)]/2\sim N_\theta (a^+)\epsilon_r(a)\sim \gamma Y\Ks^{-1/2}\ll\gamma Y$ since $N_\theta (a^+) \sim Y(\gamma/\Ks)^{1/2}\ll N_r(a)\sim Y\gamma^{1/2}$.

\bibliographystyle{elsarticle-names}
\bibliography{PMF.bib}
\end{document}